\documentclass[12pt]{article}

\usepackage{amsmath}

\usepackage{newtxtext,newtxmath}

\usepackage{graphicx}
\usepackage{here}
\usepackage{url}
\usepackage{array}
\usepackage{makecell}
\usepackage{booktabs}

\usepackage[letterpaper,margin=1in]{geometry}

\renewenvironment{abstract}
	{\quotation}
	{\endquotation}

\date{}

\makeatletter
\renewcommand{\fnum@figure}{\textbf{Figure \thefigure}}
\renewcommand{\fnum@table}{\textbf{Table \thetable}}
\makeatother

\usepackage{scicite}

\def\scititle{
	Digital Twin-Enabled Sim-to-Real Reinforcement Learning for Linear and Multi-Level Analog Memristor Programming
}
\title{\bfseries \boldmath \scititle}

\author{
	Zhuo~Diao$^{1\ast}$,
	Tetsuya~Tohei$^{1}$,
	Akira~Sakai$^{1,2}$\and
	\small$^{1}$Graduate School of Engineering Science, The University of Osaka,\\
	\small 1-3 Machikaneyama, Toyonaka, Osaka 560-8531, Japan.\and
	\small$^{2}$Spintronics Research Network Division, Institute for Open and Transdisciplinary\\
	\small Research Initiatives, The University of Osaka, 1-3 Machikaneyama, Toyonaka, Osaka 560-8531, Japan.\and
	\small$^\ast$Corresponding author. Email: diao.zhuo.es@osaka-u.ac.jp
}

\begin{document}

% Insert the title and author list
\maketitle

% Abstract, in bold
\begin{abstract} \bfseries \boldmath
Developing effective programming strategies for analog memristors remains a fundamental bottleneck for analog in-memory computing because optimization relies on costly, experiment-driven trial-and-error exploration.
This reliance arises from the nonlinear and stochastic switching dynamics of memristors, which make precise multi-level conductance programming difficult to control.
Here, we present a physics-informed digital twin framework that transforms this characterize--test--optimize development cycle into an integrated workflow, demonstrated on a four-terminal planar TiO$_{2-x}$ memristor.
The framework calibrates a physics-based model using experimental characterization data to construct a digital twin for large-scale reinforcement learning pretraining; the agent is subsequently fine-tuned on physical hardware through limited on-device interaction and deployed for multi-level programming.
Because simulated programming episodes run more than four orders of magnitude faster than on-device episodes in wall-clock time, most trial-and-error exploration is shifted from the physical device to the digital twin.
The framework reliably programs linearly and precisely spaced conductance levels and automatically discovers programming policies exhibiting different accuracy--efficiency trade-offs. The most efficient policy achieves 32-level programming using only 45 write pulses on average, approximately 13$\times$ fewer than a calibrated feedforward-pulse scheme.
Crossbar inference simulations on four image-classification benchmarks, based on the measured conductance distributions, reveal that programming accuracy requirements are strongly task-dependent: simpler tasks tolerate lower placement precision, while more demanding tasks benefit measurably from tighter programming.
These results show that reinforcement learning on a physics-informed digital twin can automatically generate programming strategies tailored to different deployment objectives, providing a practical foundation for autonomous memristor optimization and analog AI hardware deployment.
\end{abstract}

\section{Introduction}

The rapid growth of artificial intelligence (AI) has placed escalating demands on computing hardware in terms of both throughput and energy efficiency~\cite{doi:10.1126/science.aba3758}.
Modern deep neural networks require massive numbers of matrix--vector multiplication (MVM) operations, and the associated data movement between spatially separated memory and processing units in conventional von~Neumann architectures has become the dominant bottleneck for both latency and power consumption~\cite{Ambrogio:2018aa}.
Analog in-memory computing (AIMC) and neuromorphic architectures address this bottleneck by encoding synaptic weights directly as device conductances, where MVM is executed in a single analog step through Ohm's and Kirchhoff's laws, eliminating the need to shuttle data between memory and compute units~\cite{Sebastian:2020aa, Aguirre:2024aa, Zhou:2024aa}.
The resulting improvements in energy efficiency and inference latency make AIMC particularly attractive for edge AI applications~\cite{Wu:2023aa, Zuo:2025aa, bulk1, 10899870}, where analog memristors stand out as promising candidate devices owing to their high integration density, CMOS process compatibility, and ability to store analog information non-volatilely in a continuously tunable conductance state.
In crossbar-based analog neural networks, inference accuracy depends directly on how precisely the conductance of each memristor is programmed to its target weight, making reliable multi-level programmability an essential prerequisite for practical AIMC deployment.

Achieving such precise conductance control is fundamentally complicated by the intrinsic physics of oxide-based memristors~\cite{Strukov:2008aa, Li:2020aa}.
Resistive switching (RS) originates from the electric-field-driven migration of dopant ions, such as oxygen vacancies, which modulate the formation and rupture of the conduction path within the switching layer.
The applied pulse parameters do not uniquely determine the resulting conductance update: it is simultaneously shaped by defect diffusion kinetics, local electric-field redistribution, and thermal fluctuations~\cite{Kim:2014aa}.
The resulting conductance update is therefore history-dependent and susceptible to fabrication-induced disorder, causing identical programming conditions to produce different outcomes across devices and cycles~\cite{Ielmini_2016, C3NR01176C}.
Consequently, unlike digital devices, whose threshold-gated switching is protected by large noise margins that latch each cell into one of two well-separated states, hiding the underlying device physics behind a digital abstraction, the analog operation of memristors exposes these switching dynamics directly.
Every intermediate conductance is a valid, non-restoring state, with no noise margin to absorb a misplaced update. Exploiting this analog functionality therefore requires an additional layer of \emph{switching-dynamics engineering}, in which the programming process must shape, rather than merely trigger, the ion drift--diffusion kinetics that set the final conductance.

Research on memristor switching-dynamics engineering has advanced along two complementary directions: physics-based modeling of the switching process, and device-level programming on real hardware.
On the modeling side, the ion-migration kinetics underlying RS have been described at multiple levels of abstraction, ranging from compact analytical models such as the linear ion-drift model~\cite{Strukov:2008aa} and threshold-based formulations like VTEAM~\cite{Kvatinsky2015VTEAM} to finite-element method~\cite{Khot:2023aa} and density-functional theory~\cite{Kousar:2021aa}.
Such models provide physical insight and predictive power, and allow the switching behavior to be explored without repeatedly stressing the hardware.
However, they are typically used as offline design tools, decoupled from the device they represent, and they capture only the generic switching physics---neglecting the device-to-device variation, cycle-to-cycle drift, and diverse deployment conditions that govern the behavior of an individual fabricated device.
On the programming side, control strategies act directly on the real device, ranging from feedforward pulse schemes~\cite{D4NH00623B} and closed-loop write-and-verify protocols~\cite{Rao:2023aa} to learning-based controllers that infer pulses or actions from measured feedback~\cite{10595913, Spetzler:2025aa, 11173161} (reviewed in Section~\ref{sec:related}).
Because these strategies are grounded in the actual device response, they can achieve high programming precision in practice.
Their limitation is that they must rely on physical characterization and trial-and-error on real hardware.
Such interaction unfolds on real and extensive experimental timescales and typically requires large volumes of data gathered through repeated device testing; moreover, their lack of physical interpretability means that this testing routine must be repeated for every new device process, substantially increasing the cost of memristor device development.
The two directions are therefore complementary: modeling breaks free of the real-device timescale, is physically structured, and is hardware-free but only approximate, whereas programming is device-accurate but experimentally costly and device-specific.
Yet few studies have been able to close the residual model--device gap.

The concept of a physics-informed digital twin---a system that couples the physical device with a calibrated simulation model---offers a principled foundation for such a framework.
A physics-based memristor model serves as a simulation environment for automated, high-throughput search over programming strategies at negligible experimental cost, while an automated hardware interface enables real-time characterization, on-device adaptation, and direct deployment of the trained agent---eliminating manual handoffs between development stages and enabling performance estimation across application objectives (such as analog neural-network inference accuracy) before committing to real-device programming.
Within this framework, a reinforcement learning (RL) agent~\cite{Silver:2016aa, Vinyals:2019aa, Tang:2025aa} is particularly well suited to serve as the vehicle that couples the simulated and physical environments.
An RL agent formulates programming as a closed-loop sequential decision problem, learning adaptive voltage-selection strategies from environment interaction and providing intrinsic online error correction.
Training such an agent, however, requires a large number of environment interactions, which makes direct training on physical hardware prohibitively expensive because each write-and-measure cycle incurs substantial real-time overhead---a bottleneck that the digital twin likewise resolves.
It enables policy pretraining in simulation at speeds orders of magnitude faster than practical operation, with subsequent real fine-tuning closing the simulation-to-reality gap through a limited number of interactions with the physical environment~\cite{Tan:2026aa, Wang:2025aa}.
While such closed-loop concepts are well established in robotics~\cite{Zhao2020,Tobin2017} and scientific discovery~\cite{Abolhasani:2023aa, Diao:2025aa, Okuyama:2025aa}, to the best of our knowledge, a framework that integrates automated device testing, simulation-based policy optimization, on-device adaptation, and application-level evaluation into a single workflow has not yet been reported.

In this work, we present a physics-informed digital twin framework for automated testing and deployment of analog memristor programming strategies, demonstrated on a four-terminal planar TiO$_{2-x}$ memristor.
The framework integrates three tightly coupled components: (i) an automated measurement interface that executes programmable control logic directly on hardware, enabling both systematic device characterization and on-device RL interaction; (ii) a generalized physics-based resistive-switching model aligned to experimental data, which serves as the digital twin environment for large-scale RL pretraining; and (iii) a curriculum-trained RL agent acting as the programming decision engine, pretrained in simulation and refined on physical hardware through limited on-device fine-tuning.
We validate the framework at both device and system levels.
At the device level, we demonstrate reliable 32-level linear conductance programming and show that curriculum learning enables automatic discovery of policies spanning distinct accuracy--efficiency trade-offs, with the most efficient policy achieving programming in only 45 write pulses on average---approximately 13$\times$ fewer than a calibrated feedforward-pulse scheme.
At the system level, crossbar inference simulations on four image-recognition benchmarks, which draw the synaptic weights from the measured conductance distributions, reveal that programming accuracy requirements are strongly task-dependent, establishing a quantitative link between memristor programming fidelity and analog neural-network inference accuracy.
These results demonstrate that the proposed digital twin framework can generate programming strategies tailored to diverse application requirements, and establish a practical methodology for physics-informed digital-twin-assisted memristor characterization and deployment toward autonomous analog AI hardware development.

This paper is organized as follows.
Section~\ref{sec:related} reviews related work on multi-level memristor programming, learning-based programming, and device models and digital twins of memristive systems.
Section~\ref{sec:method} describes the measurement system, the physics-based device model, and the RL framework.
Section~\ref{sec:results} presents the calibration of the digital twin, sim-to-real training, an analysis of the learned policies, a comparison of curriculum designs, and the application-level evaluation.
Section~\ref{sec:limitations} discusses the limitations of this study and directions for future research, and Section~\ref{sec:conclusion} concludes the paper.

\section{Related Works}\label{sec:related}

Programming memristors to desired multi-level analog states has been explored through a wide range of approaches that balance programming speed, accuracy, and adaptability to device variability.
One representative class is the self-terminating approach, which employs analog feedback circuitry to terminate programming once the target conductance is reached~\cite{7727298}.
This strategy enables rapid programming with low latency; however, its precision is limited by the analog feedback circuitry, typically restricting the conductance resolution to only 4--8 program levels~\cite{Wan:2022aa, 10.3389/fnano.2026.1866642}.
Alternatively, open-loop programming improves conductance controllability by applying non-identical write pulses rather than fixed programming voltages.
By gradually increasing the write-pulse amplitude as the pulse sequence progresses, it enables the modulation of 32 programmable states in ferroelectric hafnium zirconium oxide (HfZrO$_x$, HZO)-based synaptic devices~\cite{7912345}, and 64 conductance levels with an average programming error below 0.5\% in one-transistor--one-resistor (1T1R) tantalum oxide arrays~\cite{Merced-Grafals_2016}.
Despite their high programming speed and hardware efficiency, these open-loop approaches fundamentally rely on prior calibration. Consequently, their accuracy is strongly limited by calibration quality. It can degrade under cycle-to-cycle and device-to-device variability, since no feedback mechanism exists to compensate for instantaneous programming errors.
Closed-loop protocols, commonly referred to as iterative write-and-verify (IWV) schemes, address device variability by interleaving read-back measurements with write pulses, applying corrective pulses until the measured conductance falls within a predefined target window. Using this strategy, previous studies have demonstrated ultra-high-precision programming with up to 2048 levels~\cite{Rao:2023aa}.
A feedforward pulse (FFP) scheme was recently proposed to reduce the number of verification operations while achieving 512 addressable linear conductance states~\cite{D4NH00623B}.
Although IWV methods can reliably achieve sub-level-spacing accuracy, the repeated read-write operations introduce substantial programming overhead.
The number of required correction iterations is inherently unpredictable, leading to time-intensive deployment that complicates latency-sensitive applications.

To balance programming speed and accuracy, recent studies have introduced machine-learning-driven approaches to optimize pulse programming strategies.
For example, a neural network has been employed to predict the pulse duration required to achieve a desired conductance change from the current device state~\cite{10595913}.
Other works model conductance evolution under different pulse amplitudes and widths, enabling predictive pulse selection for analog programming~\cite{Spetzler:2025aa}.
These learning-based approaches can significantly reduce the number of programming iterations; however, they remain fundamentally open-loop during inference. As a result, their performance is still susceptible to device variation errors. In addition, collecting sufficiently large training datasets for each device type is often a time-consuming process, increasing both characterization complexity and device development cost.
RL is a standard framework for solving sequential decision-making problems formulated as Markov decision processes (MDP)~\cite{sutton2018reinforcement, Valverde:2023aa}.
Analog memristor programming naturally exhibits the characteristics of an MDP: each programming pulse modifies the device conductance, thereby changing the system state and influencing the effectiveness of subsequent programming actions.
The objective is therefore to learn a policy that adaptively selects the programming voltage according to the current device state so as to reach the target conductance with minimum programming cost.
Finally, the RL agent can be naturally integrated into a closed-loop programming system, enabling continuous interaction with the device and intrinsic online error correction through state-feedback.

Relative to these studies, the present work advances device programming toward finer and more reliable control by addressing sequential linear multi-level programming, in which the agent must traverse all target resistance levels without overshoot.
It further broadens the evaluation from conventional device-level programming metrics to application-level inference accuracy, thereby establishing a direct connection between programming performance and downstream computing capability.
At the system level, the same agent and environment interface are deployed consistently in simulation and hardware, enabling a unified sim-to-real pipeline that integrates simulation pretraining, on-device fine-tuning, and deployment.
In addition, curriculum-based reward scheduling is introduced to learn policies operating at different accuracy--efficiency trade-offs. Finally, by examining the learned policies throughout the development pipeline, the present work quantifies how voltage-regime-dependent inaccuracies of the digital twin contribute to the sim-to-real performance gap and reveals how programming precision propagates to task-dependent inference accuracy.

\section{Method}\label{sec:method}

\subsection{System architecture overview}

\begin{figure}[htbp]
	\centering
	\includegraphics[width=.7\textwidth]{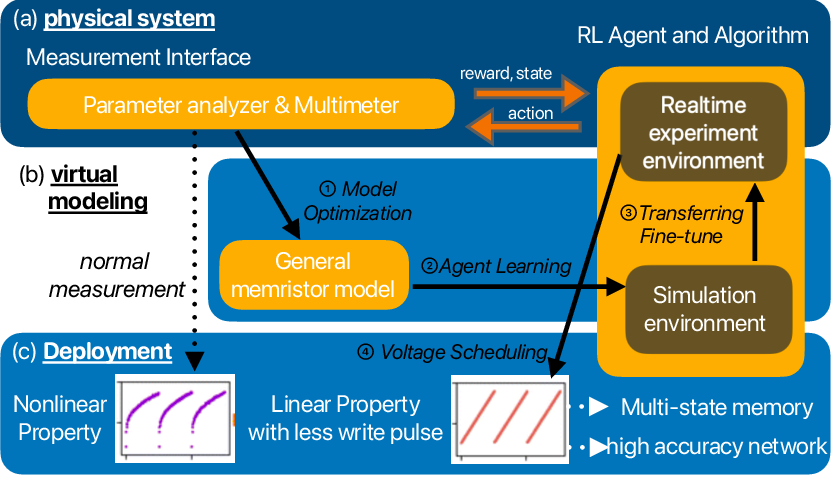}
	\caption{
Schematic of the digital twin system for memristor programming. (a) Physical measurement interface for device characterization, control, and online RL execution. (b) Virtual modeling environment for simulation-based RL training using a fitted physics-based generalized memristor model. (c) Deployment of the trained RL agent on the physical platform, where it serves as a memristor controller for multi-level memory and neural network applications.
	}
	\label{FIG:framework}
\end{figure}

Fig.~\ref{FIG:framework} presents an overview of the proposed sim-to-real RL framework, which is organized into three layers.
The \textit{physical system} layer (Fig.~\ref{FIG:framework}a) encapsulates external instrument-control application programming interfaces (APIs) and enables automated memristor characterization and programming. It supports device selection through an integrated electronic switching matrix, write-pulse application, and resistance readout. This layer also serves as a standardized Gymnasium-compatible environment, allowing on-device RL during hardware fine-tuning and evaluation. By integrating experiment execution, data acquisition, and analysis, the framework establishes a closed-loop workflow that autonomously performs measurement, decision-making, and control.
The \textit{virtual modeling} layer (Fig.~\ref{FIG:framework}b) constructs a physics-based simulation environment using analytical memristor models.
Device-model parameters are identified from experimental data acquired through the \textit{physical system} layer, enabling the creation of a computationally efficient digital-twin environment for training device-control policies.
To bridge the simulation-to-reality gap, the pre-trained RL agent is subsequently fine-tuned on the physical hardware, allowing adaptation to device characteristics that are not fully captured by the simulation model.
After training is completed, the learned policy can be deployed in the \textit{deployment} layer (Fig.~\ref{FIG:framework}c). The objective of this layer is to achieve high-precision and high-efficiency memristor programming, thereby enabling practical applications such as multi-level memory storage, in-memory computing, and analog weight loading for neural network inference.

\subsection{Measurement system}\label{sec:measurement}

The measurement system consists of two primary instruments: a Keithley 3706A switch/multimeter system and a semiconductor parameter analyzer, either a Keysight B1500A or a Keysight B2900 series source-measure unit. All measurements were performed at room temperature under ambient air conditions. The Keithley 3706A switching matrix routes electrical connections to the designated device channel, after which the semiconductor analyzer applies voltage pulses and acquires the resulting current or resistance response.
Instrument control and data acquisition are managed by a Python-based automation platform that communicates with all instruments over the Virtual Instrument Software Architecture (VISA) protocol. Standard Commands for Programmable Instruments (SCPI) are used to operate the Keysight B1500A and B2900, while Test Script Processor (TSP) commands are used for the Keithley 3706A.
The platform is built on top of PyMeasure and is openly available as an open-source repository~\cite{MemDigitalTwin}.
The RL agent is implemented using the stable-baselines3 Python library~\cite{stable-baselines3}.

\subsection{Device physics modeling}

\subsubsection{Memristor device}\label{sec:device}

The device used in this study is a four-terminal planar TiO$_{2-x}$ memristor. Unlike conventional two-terminal memristors, the four-terminal architecture enables electric fields to be applied from multiple directions to the resistive switching layer, allowing two-dimensional control of oxygen-vacancy migration and distribution within the device plane. This additional degree of freedom improves the controllability of the resistive switching process and enhances the reproducibility of the device response to programming pulses, thereby supporting more precise conductance tuning (see Appendix~\ref{sec:4t} for details).
Using the programming scheme described in Appendix~\ref{sec:4t}, the device exhibits bipolar resistive switching behavior. Positive write voltages induce potentiation, corresponding to an increase in conductance, whereas negative write voltages produce depression, resulting in a decrease in conductance. Owing to the improved switching uniformity provided by the four-terminal design, the device can be programmed to multiple intermediate conductance states with high repeatability, making it a suitable platform for evaluating precision memristor control strategies.

To characterize the switching dynamics, we applied write pulses of twelve amplitudes ($|V|$ = 0.3, 0.5, 0.75, 1.0, 1.5, 2.0, 2.5, 3.0, 4.0, 5.0, 6.0, 7.0~V) in both polarity directions.
Each write pulse had a duration of 2~s.
For each voltage amplitude, the same pulse was repeatedly applied for 200 cycles, and the device resistance was measured after every pulse. This procedure generated a set of resistance-evolution trajectories as a function of pulse number, $n_{\mathrm{pulse}}$, under a fixed programming voltage $V_t$.
The resulting voltage-dependent switching dataset was subsequently used for fitting the physics-based memristor model. Specifically, the model describes the temporal evolution of the device resistance as a function of the programming voltage and pulse count, $R = f(V_t, n_{\mathrm{pulse}})$, thereby capturing the conductance update dynamics of the memristor under repeated pulse stimulation.

\subsubsection{General physics model}

We adopt the Voltage Threshold Adaptive Memristor (VTEAM) model~\cite{Kvatinsky2015VTEAM} as the physical modeling framework. The VTEAM model is derived from the seminal memristor model proposed by Strukov~\cite{Strukov:2008aa}, which describes resistive switching through an internal state variable $x$ representing the boundary position between the doped and undoped regions within the active switching layer of thickness $d$. The normalized state variable, $\hat{x}=x/d \in [0,1]$, is assumed to vary monotonically with the device resistance and is related to the resistance through a linear interpolation between the limiting low and high resistances of the model, $R_\mathrm{on}$ and $R_\mathrm{off}$:

\begin{equation}
    R(\hat{x}) = R_\mathrm{on}\,\hat{x} + R_\mathrm{off}\,(1 - \hat{x}),
    \label{eq:vteam_resistance}
\end{equation}

\noindent and the conductance satisfies $G = 1/R$. The corresponding instantaneous current is $I = V/R(\hat{x})$. The state dynamics are governed by a voltage-driven equation:

\begin{equation}
    \frac{dx}{dt} =
    \begin{cases}
        k_\mathrm{off} \left(\dfrac{V}{V_\mathrm{off}} - 1\right)^{\alpha_\mathrm{off}} f(\hat{x}), & V \geq V_\mathrm{off} > 0, \\[8pt]
        k_\mathrm{on}  \left(\dfrac{V}{V_\mathrm{on}}  - 1\right)^{\alpha_\mathrm{on}}  f(1-\hat{x}), & V \leq V_\mathrm{on}  < 0, \\[8pt]
        0, & \text{otherwise,}
    \end{cases}
    \label{eq:vteam_dxdt}
\end{equation}

\noindent where $k_\mathrm{on}$ and $k_\mathrm{off}$ are rate constants, $\alpha_\mathrm{on}$ and $\alpha_\mathrm{off}$ are nonlinearity exponents, and $V_\mathrm{on}<0<V_\mathrm{off}$ denote the switching threshold voltages. The function $f(\cdot)$ is a window function that constrains the state evolution near the boundaries of the state space.

To better capture the switching characteristics of the four-terminal planar TiO$_{2-x}$ memristor, we extend the standard VTEAM formulation by modifying the window function.
We introduce a generalized window function $f(\hat{x},V)$ to improve the representation of the experimentally observed voltage-dependent switching behavior.
Among several candidate forms, the voltage-driven exponential function achieved the highest fitting accuracy and was therefore selected for the device model.

\begin{equation}
    f(\hat{x}, V) = \exp\!\left(-\gamma\,\hat{x} + h\,|V|\right),
    \label{eq:window}
\end{equation}

\noindent where $\gamma>0$ controls the suppression of state evolution as the device approaches the switching boundaries, while $h\ge0$ introduces a voltage-dependent factor that allows larger programming voltages to accelerate vacancy transport. For potentiation and depression processes, separate parameter pairs $\{\gamma_\mathrm{on},h_\mathrm{on}\}$ and $\{\gamma_\mathrm{off},h_\mathrm{off}\}$ are used to account for asymmetric switching behavior.

The boundary resistance values $R_\mathrm{on}$ and $R_\mathrm{off}$ are fixed at 900~$\Omega$ and 1400~$\Omega$, respectively. These values bracket the full resistance range observed during characterization at a maximum of 7~V within approximately 1000--1200~$\Omega$ (Figs.~\ref{FIG:vteam}a and \ref{FIG:vteam}b), while also encompassing the wider operating range that can potentially be accessed under higher programming voltages.
The remaining parameters---$k$, $\alpha$, and $V_\mathrm{th}$ in Eq.~(\ref{eq:vteam_dxdt}), and $\gamma$ and $h$ in Eq.~(\ref{eq:window})---are treated as free parameters and optimized to minimize the discrepancy between simulated and experimentally measured resistance trajectories across all tested programming voltages. The goal of this fitting procedure is not to extract physically unique material parameters, but to obtain a compact, computationally efficient surrogate model that reproduces the observed switching dynamics within the operating window and can serve as a high-fidelity simulation environment for RL training.

\subsubsection{Model Optimization}

The free parameters of the device model, $\boldsymbol{\theta}=(k, \alpha, V_\mathrm{th}, \gamma, h)$, are estimated separately for the potentiation and depression directions by simultaneously fitting the simulated resistance trajectories to all twelve experimental voltage conditions.
The fitting objective is the weighted mean-squared error:

\begin{equation}
    \hat{\boldsymbol{\theta}} = \arg\min_{\boldsymbol{\theta}} \sum_{k=1}^{K} w_k \sum_{n=1}^{N_k} \bigl(R_k^{(n)}(\boldsymbol{\theta}) - \tilde{R}_k^{(n)}\bigr)^2,
    \label{eq:fitting_objective}
\end{equation}

\noindent where $k$ indexes the voltage condition ($K=12$), $N_k$ is the number of applied pulses at condition $k$, $R_k^{(n)}(\boldsymbol{\theta})$ is the model-predicted resistance after the $n$-th pulse, $\tilde{R}_k^{(n)}$ is the corresponding experimental measurement, and $w_k$ is a per-condition weighting factor. Larger values of $w_k$ increase the contribution of a given voltage condition to the overall fitting objective and therefore encourage the optimizer to reproduce its switching trajectory more accurately.
For a given $\boldsymbol{\theta}$, the model is numerically integrated over the applied pulse sequence to produce a resistance trajectory, and the error is summed across all pulse indices and voltage conditions. Fitting all twelve voltage amplitudes simultaneously constrains the model to reproduce the global switching behavior across the full action space and prevents over-specialization to any single programming regime.
To improve the fidelity of the model at the boundaries of the programming range, the lowest- and highest-voltage conditions ($|V|=0.3$~V and $|V|=7.0$~V) are assigned $w_k=2$, while all intermediate voltage conditions are assigned $w_k=1$. These boundary conditions define the minimum and maximum conductance-update regimes of the device and therefore play a critical role in constraining the simulator over the full programming range.

The objective function in Eq.~(\ref{eq:fitting_objective}) is highly non-convex owing to the threshold nonlinearities and the exponential window function, making gradient-based optimization susceptible to local minima. We therefore employ Differential Evolution~\cite{Storn:1997aa}, a population-based global search algorithm that explores the parameter space without requiring gradient information and is well-suited to this class of non-smooth, multimodal landscapes.

\subsection{Implementation of RL framework}\label{sec:rl}

\begin{figure}[htbp]
	\centering
	\includegraphics[width=.85\textwidth]{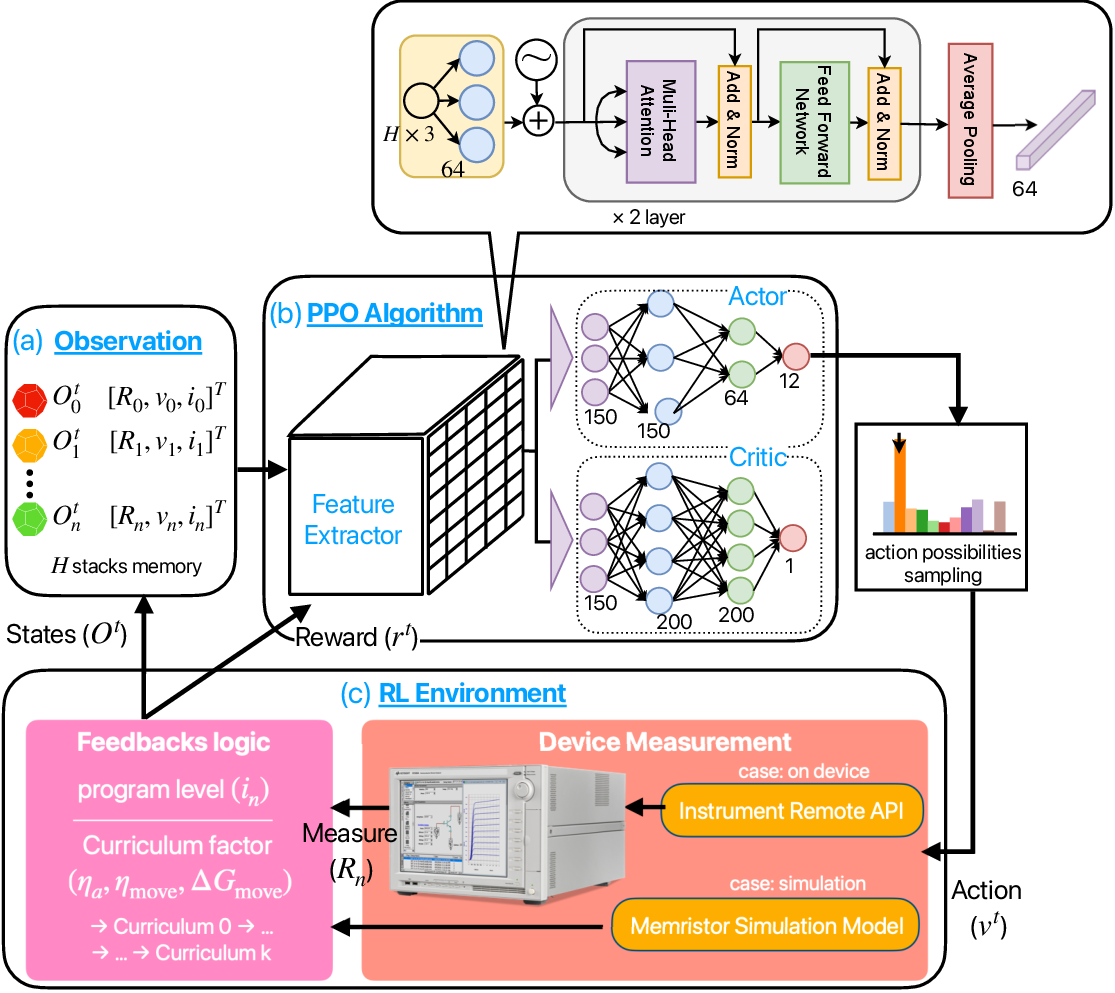}
	\caption{Architecture of the reinforcement learning (RL) framework. (a) Observation structure. The memory state $O_n^t$ at the $n$th programming step contains the measured resistance $R_n$, the applied write voltage $v_n$, and the normalized program-level index $i_n$. (b) PPO algorithm with a Transformer-based backbone network. (c) RL environment for memristor programming and state transition evaluation.}
	\label{FIG:rl_net}
\end{figure}

Fig.~\ref{FIG:rl_net} summarizes the RL framework used for memristor programming.
The agent observes a short history of device measurements (Fig.~\ref{FIG:rl_net}a; Section~\ref{sec:obs}), selects the next write voltage using a PPO policy with a Transformer-based feature extractor (Fig.~\ref{FIG:rl_net}b; Section~\ref{sec:algo}), and interacts with an environment that is either the digital twin or the physical device (Fig.~\ref{FIG:rl_net}c; Section~\ref{sec:env}).
The reward function that guides the learning process is described in Section~\ref{sec:reward}.

\subsubsection{Observation and action space}\label{sec:obs}

The memristor programming task is formulated as a sequential decision-making problem:
The target conductance range between the experimentally determined HRS and LRS is linearly and uniformly divided into $L$ target levels. Starting from the HRS, the agent aims to sequentially reach each target level while minimizing placement error and avoiding level skipping.
At each programming step $n$ within episode $t$, the agent receives a three-dimensional observation vector $\mathbf{O}_n^t$ as illustrated in Fig.~\ref{FIG:rl_net}a:

\begin{equation}
    \mathbf{O}_n^t = \left[\, R_n,\; v_n,\; i_n \,\right]^T,
    \label{eq:observation}
\end{equation}

\noindent where $R_n \in [0,1]$ is the normalized measured resistance (equivalently, the normalized conductance $\bar{G}_n$), $v_n$ is the write voltage applied at step $n$, and $i_n = \ell_n / L$ is the normalized program-level index with $\ell_n \in \{0,1,\dots,L\}$ indicating how many of the $L$ target conductance levels have been successfully reached.
Because memristive switching is path-dependent, a single observation $\mathbf{O}_n^t$ is insufficient to characterize the device state.
The agent therefore maintains a sliding window of the last $H$ steps, forming a stacked state matrix $s_t = [\mathbf{O}_{n-H+1}^t, \dots, \mathbf{O}_n^t] \in \mathbb{R}^{H \times 3}$, as depicted by the $n$-stack memory in Fig.~\ref{FIG:rl_net}a. The history buffer is padded at the start of each episode.

The action space $\mathcal{A}^t$ is discrete: $\{1, 2, \dots, 12\}$, each index mapping to one of twelve write voltages.
At programming step $n$, selecting action $a_n \in \mathcal{A}^t$ applies the corresponding write voltage $v^t_n$ to the device. After the programming pulse train is completed, the resulting conductance (or resistance) is measured and incorporated into the next observation $\mathbf{O}_{n+1}^t$.

\subsubsection{RL algorithm}\label{sec:algo}

The agent is trained using Proximal Policy Optimization (PPO)~\cite{PPO}, an on-policy actor-critic algorithm. The standard clipped surrogate objective is adopted:

\begin{equation}
    \mathcal{L}^\mathrm{CLIP}(\phi) = \mathbb{E}_t\!\left[\min\!\left(\rho_t(\phi)\,\hat{A}_t,\; \mathrm{clip}\!\left(\rho_t(\phi), 1-\epsilon, 1+\epsilon\right)\hat{A}_t\right)\right],
    \label{eq:ppo_clip}
\end{equation}

\noindent where $\phi$ denotes the policy network parameters; $s_t = [\mathbf{O}_{n-H+1}^t, \dots, \mathbf{O}_n^t] \in \mathbb{R}^{H \times 3}$ is the stacked observation matrix serving as the state input to the policy; $\rho_t(\phi) = \pi_\phi(a_t \mid s_t) /\pi_{\phi_\mathrm{old}}(a_t \mid s_t)$ is the importance-sampling ratio between the current policy $\pi_\phi$ and the policy $\pi_{\phi_\mathrm{old}}$ used to collect the rollout data; $\hat{A}_t$ is the generalized advantage estimate~\cite{schulman2018}, which measures how much better action $a_t$ is relative to the average behavior at state $s_t$; and $\epsilon$ is a small clipping threshold that constrains the policy update step by preventing $\rho_t(\phi)$ from deviating too far from 1, thereby avoiding destructively large gradient updates. A value-function loss and an entropy bonus are added to improve the accuracy of value estimation and to maintain sufficient exploration, respectively.

Fig.~\ref{FIG:rl_net}b illustrates the PPO-based policy architecture used for the memristor programming task. Because memristive switching exhibits strong history dependence, the conductance response to a given programming pulse is influenced not only by the current conductance state but also by the sequence of previously applied pulses. Consequently, the policy network must infer the effective device state from a short history of observations rather than from a single measurement.
To capture these temporal dependencies, the stacked observation sequence $s_t$ is first projected into a 64-dimensional embedding space through a linear layer, after which a learnable positional encoding is added to preserve the ordering of observations within the history window. A feature extractor then processes the resulting sequence.
The output tokens are averaged across the history dimension to generate a 64-dimensional latent feature vector that summarizes the recent programming trajectory.
The actor and critic share this common feature and subsequently branch into two independent output heads. The \textit{actor} maps the latent feature vector to a probability distribution $\pi_\phi(a_t \mid s_t)$ over the 12 discrete write-voltage actions, from which the next programming action is sampled. The \textit{critic} receives the same feature representation and outputs a scalar state-value estimate $v^t(\phi(s_t))$, which is used to compute the generalized advantage estimate $\hat{A}_t$ for policy optimization. This shared-backbone architecture enables efficient feature learning while allowing the policy and value functions to specialize in action selection and value estimation, respectively.

Among the evaluated feature-extractor architectures, a Transformer-based backbone achieved the fastest reward convergence and the highest final performance. The adopted feature extractor consists of two stacked Transformer encoder layers, each comprising a multi-head self-attention module followed by a feed-forward network, with residual connections and layer normalization applied after each sub-layer. Ablation studies further indicate that a history length of $H=4$ observations provides the best trade-off between programming accuracy and training efficiency. Increasing the history length beyond this value yields negligible performance gains while increasing computational cost.
As for the observation-history buffer, compared with conventional zero padding, filling the unavailable history entries using the earliest available observation results in higher programming accuracy and more stable policy behavior.
A detailed analysis of network architectures, history lengths, and padding strategies is provided in Appendix~\ref{sec:nn}.

\subsubsection{RL environment}\label{sec:env}

As illustrated in Fig.~\ref{FIG:rl_net}c, the RL framework supports two interchangeable environment modes: a simulation mode and a hardware mode. Both modes share identical observation structures, action definitions, reward functions, target-level specifications, and termination conditions. This interface consistency is the key enabler of sim-to-real transfer: a policy trained entirely in simulation can be deployed on the physical device without any architectural modification.

In \textit{simulation mode}, the selected write voltage $v_t$ is applied to the fitted VTEAM model. At each programming step, the model integrates the switching dynamics to update the internal state variable $\hat{x}$, and the device resistance is computed via Eq.~(\ref{eq:vteam_resistance}). Additive uniform noise is applied to the simulated resistance to model measurement uncertainty and cycle-to-cycle variability.
In \textit{hardware mode}, the same action $v_t$ is issued to the physical instrument through the remote-control API (Section~\ref{sec:measurement}), which applies the corresponding voltage pulse to the memristor and reads back the resulting resistance $R_n$ directly from the device.

Each episode begins with the device initialized to the HRS. The agent's goal is to drive the conductance through $L$ linearly spaced target levels from $G_\mathrm{HRS}$ to $G_\mathrm{LRS}$, defined as:

\begin{equation}
    G^*_\ell = G_\mathrm{HRS} + \ell \cdot  dG, \quad  dG = \frac{G_\mathrm{LRS} - G_\mathrm{HRS}}{L},
    \label{eq:target_levels}
\end{equation}

\noindent where $\ell \in \{1, \dots, L\}$ indexes the target levels.
After each programming step, the environment measures the device resistance $R_n$ and passes it to the Feedback logic (Fig.~\ref{FIG:rl_net}c, left), which computes the updated program-level index $i_t=\ell_n/L$.
A level transition from $\ell_n$ to $\ell_n+1$ is registered when the measured conductance $G_n$ falls within a tolerance window $\delta = \ell_\mathrm{tol} \cdot  dG$ centered on the next target $G^*_{\ell_n+1}$, where $\ell_\mathrm{tol}$ is the tolerance factor. Tighter values of $\ell_\mathrm{tol}$ enforce higher placement accuracy but increase the risk of overshoot. The episode terminates when all $L$ levels have been reached, or the maximum step budget is exhausted.
The Feedback logic also assembles the next observation $\mathbf{O}^t$ and evaluates the step reward $r^t$.

\subsubsection{Reward function}\label{sec:reward}

The reward function is designed to balance five objectives:
(i) minimization of the total number of programming steps,
(ii) accurate level placement,
(iii) efficient conductance progression,
(iv) avoidance of level overshoot, and
(v) complete traversal of the target conductance range.

The reward $r_n$ at step $n$ is:

\begin{equation}
    r_n = r_n^\mathrm{step} + r_n^\mathrm{level} + r_n^\mathrm{move} + r_n^\mathrm{jump} + r_n^\mathrm{term}
    \label{eq:reward_total}
\end{equation}

\noindent \textbf{Step penalty.} A fixed cost is subtracted at every programming step:

\begin{equation}
    r_n^\mathrm{step} = -\eta_\mathrm{step},
    \label{eq:reward_step}
\end{equation}

\noindent where $\eta_\mathrm{step} \geq 0$ is a constant per-step penalty. It creates pressure to complete the programming task in as few steps as possible.

\noindent \textbf{Level success reward.} When the agent transitions the device to the next target level (i.e., $\ell_n = \ell_{n-1} + 1$), it receives a positive reward that decays exponentially with the placement error $\delta_n = |G_n - G^*_{\ell_n}|$:

\begin{equation}
    r_n^\mathrm{level} = \exp(-\eta_a \cdot \delta_n) \cdot \mathbf{1}[\text{level success}],
    \label{eq:reward_level}
\end{equation}

\noindent where $\eta_a \geq 0$ is the distance penalty factor that is gradually increased across curriculum stages to encourage precise placement.

\noindent \textbf{Movement penalty.} At steps without a level transition, the agent is penalized if the applied pulse failed to produce a sufficient conductance increment, $\Delta G_\mathrm{move}$.

\begin{equation}
    r_n^\mathrm{move} = \begin{cases}
        0 & \Delta G_n \geq \Delta G_\mathrm{move}, \\
        -\eta_\mathrm{move}, & |\Delta G_n| < \Delta G_\mathrm{move} \\
        -2\eta_\mathrm{move}, & \Delta G_n < -\Delta G_\mathrm{move}\cdot 0.1,
    \end{cases}
    \label{eq:reward_move}
\end{equation}

\noindent where $\Delta G_n = G_n - G_{n-1}$ is the conductance change and $\eta_\mathrm{move}$ is the movement-failure penalty coefficient. By penalizing programming actions that fail to generate a conductance increment larger than $\Delta G_\mathrm{move}$, this term discourages inefficient voltage selections and encourages rapid progression through the conductance range. As a result, the learned policy tends to complete the programming task using fewer pulse steps.

\noindent \textbf{Jump penalty.} If the agent skips one or more levels ($\ell_n > \ell_{n-1} + 1$), a penalty proportional to the number of skipped levels is applied:

\begin{equation}
    r_n^\mathrm{jump} = -\mu_\mathrm{jump} \cdot \max(0, \ell_n - \ell_{n-1} - 1),
    \label{eq:reward_jump}
\end{equation}

\noindent where $\mu_\mathrm{jump}=3$ is the jump-penalty coefficient.
A relatively strong jump penalty is employed because conductance overshoot directly degrades multi-level programming accuracy and may require additional corrective pulses. This term encourages monotonic progression through adjacent target levels and suppresses large conductance jumps that reduce the achievable programming resolution.

\noindent \textbf{Terminal penalty.} At the end of an episode, if the agent has not reached all $L$ levels, a penalty proportional to the number of unvisited levels is incurred:

\begin{equation}
    r_n^\mathrm{term} = -\mu_\mathrm{term} \cdot (L - \ell_n) \cdot \mathbf{1}[\text{done}],
    \label{eq:reward_term}
\end{equation}

\noindent where $\mu_\mathrm{term}=0.5$ is the range-failure penalty. This term encourages the agent to explore the full conductance range and prevents policies that optimize local rewards while failing to complete the overall programming task.

The reward parameters $(\eta_\mathrm{step}, \eta_a, \eta_\mathrm{move}, \Delta G_\mathrm{move})$ are scheduled through curriculum learning~\cite{cl}. Early stages emphasize successful conductance-range traversal, whereas later stages progressively prioritize accurate level placement and programming efficiency. By adjusting the relative importance of these objectives during training, the framework can produce policies optimized for either high-precision multi-level programming or reduced programming latency.

\section{Experiment and results}\label{sec:results}

\subsection{Digital twin modeling}\label{sec:dt_model}

\begin{figure}[htbp]
	\centering
	\includegraphics[width=.95\textwidth]{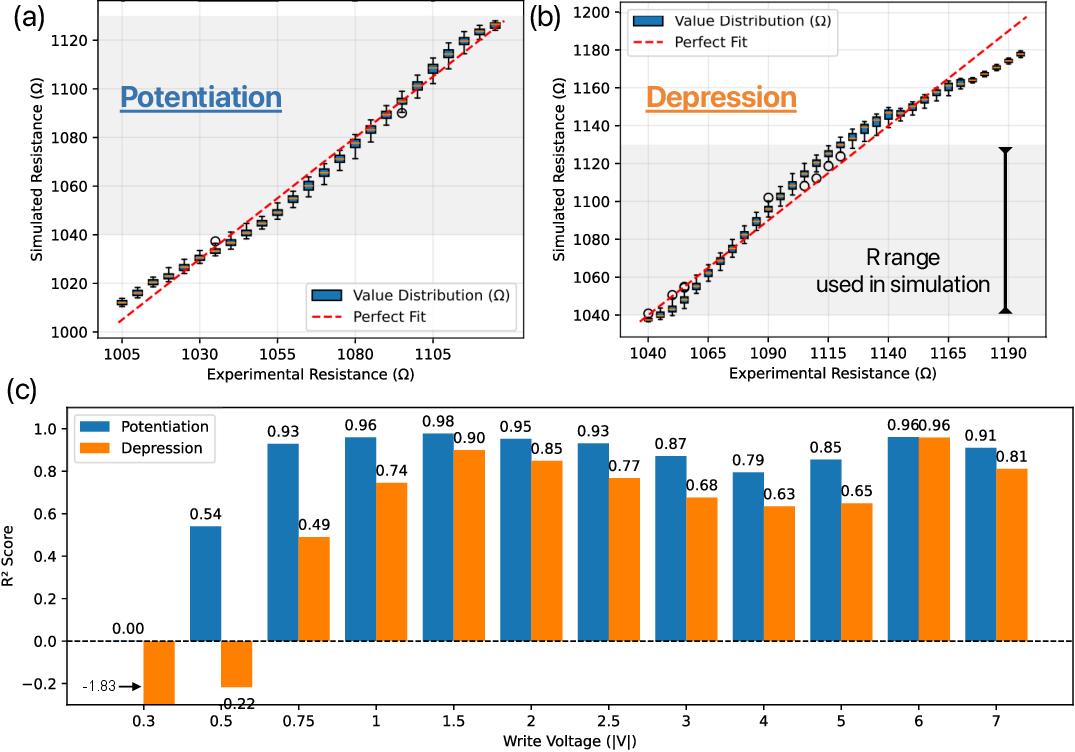}
	\caption{
VTEAM model~\cite{Kvatinsky2015VTEAM} fitting results for the 4-terminal TiO$_{2-x}$ memristor under 12 write voltage amplitudes ranging from 0.3~V to 7~V: (a) potentiation induced by positive voltages and (b) depression induced by negative voltages. The error box plots show the distribution of residuals between the model-predicted and experimentally measured resistance values across pulse indices. The gray-shaded region (1040--1130~$\Omega$) indicates the operating window used for programming.
(c) $R^2$ scores of the fitted VTEAM model evaluated separately for potentiation and depression pulse sequences across all tested write voltage amplitudes. Negative $R^2$ values indicate voltage regimes in which the model fails to accurately reproduce the measured switching dynamics.
}
	\label{FIG:vteam}
\end{figure}

Fig.~\ref{FIG:vteam} presents the results of fitting the experimentally measured temporal voltage--resistance ($V$--$R$) data of our four-terminal planar TiO$_{2-x}$ memristor device to the analytical physical model.
Figs.~\ref{FIG:vteam}a and \ref{FIG:vteam}b show boxplots comparing the experimentally measured resistance values with those predicted by the physical model during potentiation and depression, respectively. In both panels, the horizontal axis represents the measured resistance, while the vertical axis corresponds to the model-predicted resistance. The red dashed line serves as a visual reference indicating perfect agreement between the experimental data and the model predictions (see also Appendix~\ref{sec:fit} for a detailed comparison of the model predictions with the experimental measurements).
Fig.~\ref{FIG:vteam}c summarizes the fitting performance for each write voltage using the coefficient of determination ($R^2$), where potentiation and depression are represented by blue and orange markers, respectively. The fitting performance is poor in the low-voltage regime ($|V| \leq 0.5$~V), with some depression conditions even exhibiting negative $R^2$ values.
This behavior can be attributed to the fact that the analytical model neglects the influence of temperature-dependent oxygen-vacancy diffusion.
During device operation, Joule heating increases the local device temperature, thereby accelerating the resistance switching dynamics~\cite{Choi:2005aa}.
Since Joule-heating effects become increasingly significant at higher write voltages but remain weak at lower voltages, the model parameters optimized for the high-voltage regime cannot accurately describe the low-voltage switching behavior, leading to degraded fitting performance in this region.
Although the overall fitting accuracy is moderate, with $R^2=0.807$ for potentiation and $R^2=0.452$ for depression, substantially improved agreement is obtained when the resistance range is restricted to 1040--1130~$\Omega$ (gray-shaded region in Figs.~\ref{FIG:vteam}a and \ref{FIG:vteam}b). Within this practically relevant operating range, the fitting accuracy increases to $R^2=0.958$ for potentiation and $R^2=0.896$ for depression, indicating that the model captures the device dynamics with acceptable accuracy.
Because the potentiation process achieves the highest modeling accuracy, the subsequent reinforcement-learning demonstrations are conducted using potentiation.

These results define the validity range of the digital twin.
The model reproduces the measured dynamics quantitatively for potentiation within the 1040--1130~$\Omega$ operating window ($R^2=0.958$) and for write amplitudes of 0.75--7.0~V, for which the per-voltage $R^2$ values for potentiation range from 0.79 to 0.98 (Fig.~\ref{FIG:vteam}c).
In contrast, its predictions at $|V| \leq 0.5$~V and, to a lesser extent, for depression should be regarded as approximate.
These voltage-regime-specific inaccuracies are directly reflected in the sim-to-real gap of the learned policy; their implications are further discussed in Section~\ref{sec:limitations}.

\subsection{Sim-to-real training}\label{sec:sim2real}

\begin{figure}[htbp]
	\centering
	\includegraphics[width=.95\textwidth]{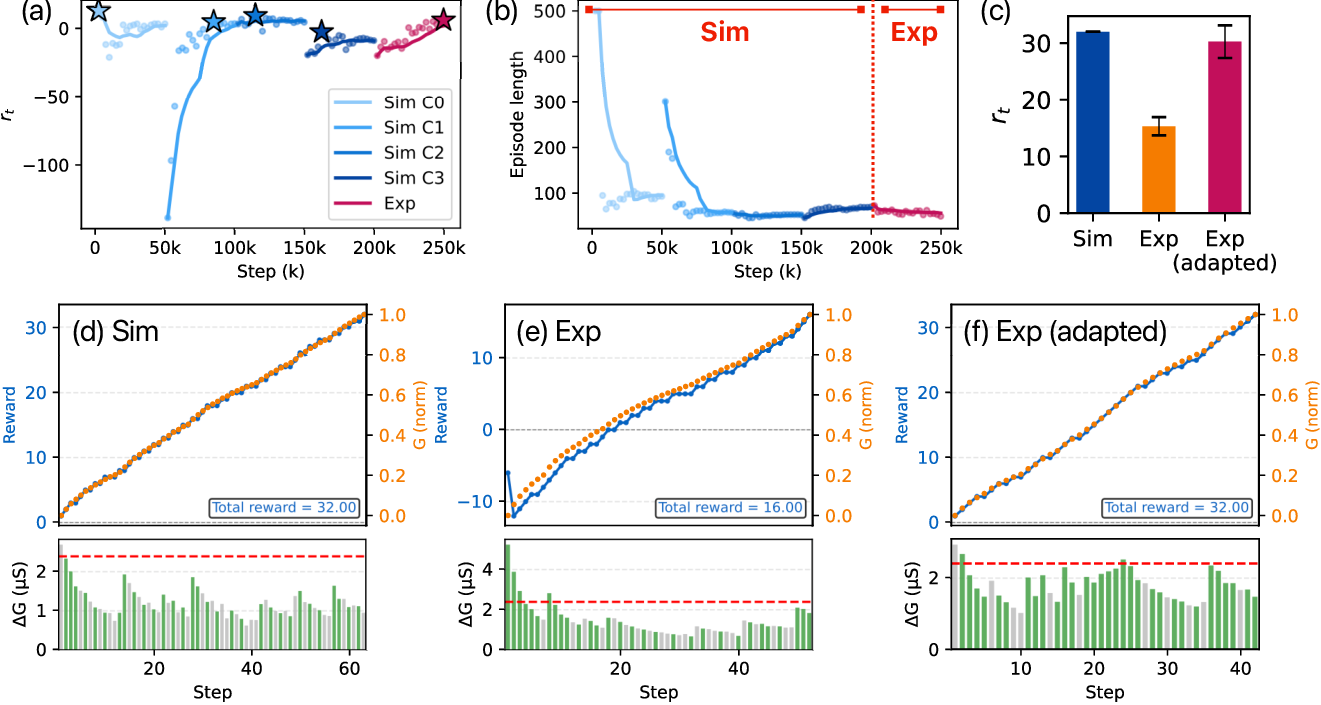}
	\caption{Sim-to-real training results. (a) Mean episodic reward, $r_t$, and (b) mean episode length during training across five sequential curriculum stages: four simulation stages (C0--C3) followed by one on-device fine-tuning stage (Exp). In the reward plot, stars indicate the best-performing checkpoint within each stage.
(c) Sim-to-real transfer performance of the curriculum learning framework. Bar charts with error bars compare the episodic reward, $r_t$, under three evaluation conditions: the final simulation environment (Sim), direct hardware deployment before fine-tuning (Exp), and hardware deployment after on-device adaptation (Exp adapted).
(d--f) Step-level episode visualization under three conditions: (d) Sim, (e) Exp, and (f) Exp adapted. Top row: cumulative reward $r_t$ (left axis, blue) and normalized conductance $G$ (right axis, orange) as functions of programming step. Bottom row: per-step conductance change $\Delta G$ ($\mu$S); green bars indicate steps that triggered a conductance-level transition, while gray bars indicate no state change. The dashed red line denotes the ideal target value of $\Delta G$.
   }
	\label{FIG:sim2real}
\end{figure}

\begin{table}[htbp]
\centering
\caption{Reward parameter schedule for Curriculum~A.}
\label{tab:curriculumA}
\begin{tabular}{llccc}
\toprule
Stage & Mode & $\ell_\mathrm{tol}$ & $(\eta_\mathrm{step},\; \eta_a,\; \eta_\mathrm{move},\; \Delta G_\mathrm{move}\;[\mu\text{S}])$ & Max steps \\
\midrule
C0  & Simulation & 0.25 & $(0.005,\; 0.00,\; 0.005,\; 0.598)$ & 500 \\
C1  & Simulation & 0.40 & $(0.500,\; 0.00,\; 0.005,\; 0.598)$ & 300 \\
C2  & Simulation & 0.40 & $(0.500,\; 0.25,\; 0.005,\; 0.598)$ & 300 \\
C3  & Simulation & 0.25 & $(0.500,\; 0.25,\; 0.005,\; 0.598)$ & 300 \\
Exp & Hardware   & 0.25 & $(0.500,\; 0.25,\; 0.005,\; 0.598)$ & 300 \\
\bottomrule
\end{tabular}
\end{table}

The RL agent is adapted to achieve linear programming across $L=32$ discrete conductance levels, with $G_\mathrm{HRS}\approx0.885$~mS and $G_\mathrm{LRS}\approx0.962$~mS by a five-stage curriculum learning strategy.
Training was first conducted in four progressively more challenging simulation environments (Sim C0--C3), allowing the agent to gradually acquire the required programming behavior. The learned policy was then transferred to the physical memristor device for on-device fine-tuning in the final stage (Exp).
The evolution of the episode reward and episode length throughout the training process is shown in Figs.~\ref{FIG:sim2real}a and \ref{FIG:sim2real}b, respectively. Clear transitions can be observed at the boundaries between curriculum stages, reflecting the increasing task difficulty and the corresponding adaptation of the agent.
The curriculum factors used in the reward function for each training stage are summarized in Table~\ref{tab:curriculumA}. During the early stages of curriculum learning, a relatively large step penalty, $\eta_\mathrm{step}$, together with a relaxed tolerance threshold, $\ell_\mathrm{tol}$, is employed to encourage efficient programming and to guide the agent toward reaching the target conductance state with a minimal number of programming operations. Starting from curriculum stage C2, the distance penalty factor $\eta_a$ is increased, forcing the agent to place greater emphasis on programming accuracy while maintaining operational efficiency.
Further details on the device performance metrics and the design rationale across the different curriculum stages are discussed in Section~\ref{sec:Curriculum}.

To quantitatively evaluate conductance overshoot across the 32 programming levels, the reward components associated with programming efficiency and action cost, namely $(\eta_\mathrm{step}, \eta_a, \eta_\mathrm{move})$, were set to zero during evaluation. Only the overshoot-related term, $\mu_\mathrm{jump}=3$, was retained for calculating the rollout reward $r_t$. Under this metric, an episode achieves the maximum score of 32 when all conductance levels are programmed without overshoot.

The Sim-to-Real transfer performance is summarized in Fig.~\ref{FIG:sim2real}c, which compares the policy performance in the simulation environment after completion of curriculum stage C3 (``Sim'' label), its direct deployment on the physical hardware (``Exp'' label), and the performance after on-device fine-tuning during the Exp stage (``Exp(adapted)'' label). While the agent achieves the maximum score in simulation, the reality gap between the fitted device model and the physical memristor significantly degrades the performance of the directly transferred policy, resulting in an average rollout reward of only 15.3. After domain adaptation through on-device fine-tuning, the average rollout reward increases substantially to 30.2, approaching the theoretical maximum.
A similar trend is observed in the overshoot statistics. The directly transferred policy exhibits an average of 0.601 conductance overshoot events per episode, whereas the fine-tuned policy reduces this value to 0.203 overshoots per episode.

To illustrate the step-level behavior underlying these aggregate statistics, we randomly selected one representative episode from each evaluation condition and visualized the evolution of the cumulative reward $r_t$ and conductance $G$ as a function of programming step (Figs.~\ref{FIG:sim2real}d--f). The bottom panel of each sub-figure shows the per-step conductance change $\Delta G$; steps that triggered a level transition are shown as green bars, steps without a transition as gray bars, and the dashed red line marks the ideal per-level $dG$ defined by Eq.~(\ref{eq:target_levels}).

In simulation (Fig.~\ref{FIG:sim2real}d), the agent consistently produces $\Delta G$ increments close to the ideal value, and the cumulative reward rises monotonically as each level is successfully reached. Upon direct deployment to hardware without fine-tuning (Fig.~\ref{FIG:sim2real}e), the behavior changes markedly during the early phase, where the conductance is still close to HRS and small write voltages are required for programming. In this regime, the agent frequently overshoots the target level, causing the cumulative reward to decrease before partially recovering at intermediate steps when higher write voltages become applicable.
This pattern is directly attributable to the low predictive accuracy ($R^2 \leq 0.54$) of the fitted VTEAM model at 0.3~V and 0.5~V (Fig.~\ref{FIG:vteam}c): the inaccurate low-voltage model trained the agent to control fine conductance steps in simulation with a dynamics that does not match real device behavior, exposing a voltage-regime-specific sim-to-real gap.
After on-device fine-tuning (Fig.~\ref{FIG:sim2real}f), the per-step $\Delta G$ distribution tightens substantially and the cumulative reward rises steadily throughout the episode, recovering the smooth progression observed in simulation. Quantitatively, the mean episode length decreases from 63.0 steps (simulation) to 45.3 steps (after fine-tuning), with 32 levels successfully reached in fewer pulses on the physical device than in the idealized simulation.

\subsection{Analysis of sim-to-real action space}\label{sec:action}

\begin{figure}[htbp]
	\centering
	\includegraphics[width=.95\textwidth]{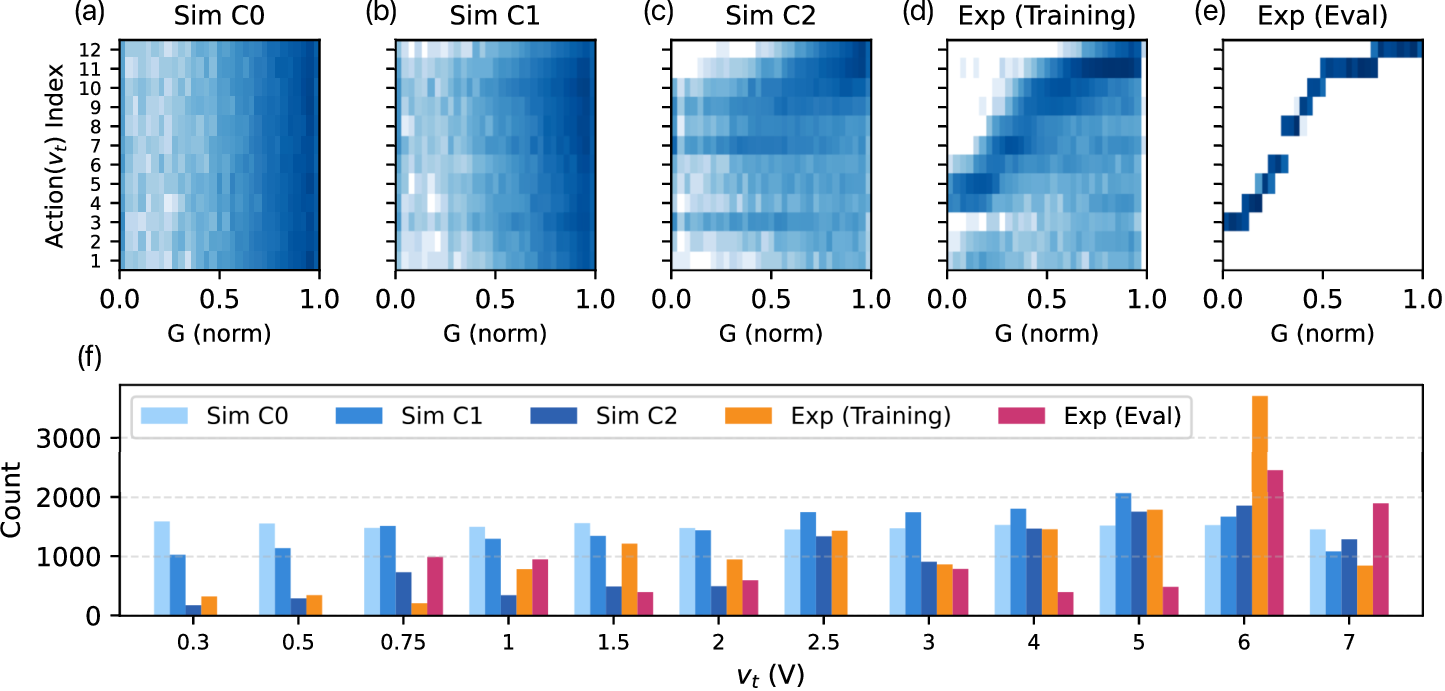}
	\caption{
Evolution of the RL policy decision space across curriculum stages. (a--e) Two-dimensional heatmaps showing the joint distribution of normalized conductance state $G$ (x-axis) and selected write-voltage action index (y-axis), accumulated over 200 episodes for each condition: simulation stages (a) C0, (b) C1, and (c) C2, followed by (d) hardware training after simulation stage C3, ``Exp (Training)'', and (e) final hardware evaluation with greedy action selection, ``Exp (Eval)''. (f) Grouped bar chart of absolute action-selection frequencies across all five conditions, revealing the progressive transition from uniformly exploratory behavior to device-physics-aware conservative voltage-selection strategies as training advances.
	}
	\label{FIG:action_space}
\end{figure}

To understand how the RL policy progressively incorporates device-physics knowledge during training, we analyze the joint distribution of the normalized conductance state $\bar{G}$ and the selected write-voltage action across the five conditions (Figs.~\ref{FIG:action_space}a--e), each accumulated over 200 episodes.
The first four conditions (Figs.~\ref{FIG:action_space}a--d) show the action distributions sampled from the policy during training. Figs.~\ref{FIG:action_space}a--c correspond to the simulation environments, whereas Fig.~\ref{FIG:action_space}d shows the policy during on-device fine-tuning, which starts from the checkpoint obtained at the end of simulation stage C3.
The final condition (Fig.~\ref{FIG:action_space}e) is obtained by greedy action selection and therefore represents the converged policy.
Higher conductance states are associated with higher write voltages, indicating that the agent has learned an effective programming strategy.

The heatmap evolution reveals two complementary trends driven by the reward function. First, the density in the upper-left region of each heatmap---corresponding to high write voltages applied at low conductance states---diminishes progressively as training advances and nearly vanishes by Exp (Training). This reflects the effect of the jump penalty $r_n^\mathrm{jump}$: at low conductance, where the device responds strongly to each pulse, large voltages risk overshooting multiple levels simultaneously, and the agent learns to avoid them. Second, the density in the lower-right region---low write voltages applied at high conductance states---also decreases, driven by both the movement penalty $r_n^\mathrm{move}$ and the physics of the device: as the conductance approaches the LRS, the window function in Eq.~(\ref{eq:window}) suppresses the state evolution, so that low voltages no longer produce the conductance increment required to reach the next level. Together, these two effects guide the policy toward a diagonal band in the heatmap, where the selected voltage increases monotonically as $\bar{G}$ increases.
Notably, the action distribution during on-device fine-tuning, which starts from the simulation-pretrained policy (Fig.~\ref{FIG:action_space}d), and that of the converged on-device policy (Fig.~\ref{FIG:action_space}e) exhibit a clear discrepancy in the low-voltage regime.
This deviation is consistent with the physics-model fitting results, which show lower goodness-of-fit ($R^2$) at 0.3 and 0.5 V, indicating that the simulator captures device dynamics less accurately under weak programming conditions.
These observations reveal a direct relationship between digital-twin fidelity and sim-to-real transfer quality: inaccuracies in the physics model are reflected in the learned control policy and subsequently require compensation through on-device adaptation. Consequently, improving model fidelity is expected to further reduce the reliance on hardware fine-tuning, highlighting accurate physics-based modeling as a key prerequisite for a robust digital twin platform in analog memristor programming.

The marginal action-selection frequencies, aggregated across all conductance states, are shown in Fig.~\ref{FIG:action_space}f. In the early simulation stages (C0 and C1), the frequency distribution over the 12 discrete voltages is nearly uniform, consistent with the broad exploratory behavior induced by the relaxed tolerance.
As training advances, the distribution becomes increasingly non-uniform: in Exp (Eval), the highest voltages (6.0 and 7.0~V), which are applied in the upper half of the conductance range (Fig.~\ref{FIG:action_space}e), are selected most frequently, whereas some actions are effectively abandoned; in particular, 0.3~V, 0.5~V, and 2.5~V are rarely selected in Exp (Eval).
This behavior demonstrates a flexibility property of the proposed framework: even when the discrete action set includes voltages that are suboptimal for a specific device, the RL agent autonomously identifies and abandons them through sufficient interaction, without requiring manual calibration of the action space.

The sim-to-real workflow offers three practical advantages that are evident from this analysis. First, physics-based simulation enables fast, parallelizable data generation and policy training.
In terms of wall-clock time, completing 200 evaluation episodes requires only 3.6--5.8~s in simulation, compared with 29--40~h on the physical device, corresponding to a speedup of more than four orders of magnitude ($>10^{4}$-fold).
Throughout this paper, this speedup refers to the wall-clock time per programming episode; it excludes the computation required for neural-network updates and does not remove the need for on-device fine-tuning (approximately $5\times10^{4}$ steps; Section~\ref{sec:sim2real}).
Because the four simulation stages account for approximately 80\% of all training steps in the present workflow, executing them in the digital twin avoids the corresponding hardware experiments. This reduction in experimental time makes large-scale policy search and curriculum optimization computationally tractable.
Second, the on-device fine-tuning stage (Exp) does not begin from a random policy but from the physics-informed checkpoint at the end of C3, whose behavior on hardware is shown in Fig.~\ref{FIG:action_space}d and which already concentrates exploration in a physically meaningful region of the action space.
This informed initialization is expected to substantially reduce the number of real-device interactions needed to close the sim-to-real gap.
Finally, the simulation environment enables inexpensive hyperparameter search, including reward coefficients, curriculum schedules, and network architecture, before any hardware experiment is conducted, eliminating the risk of device degradation during policy search.

\subsection{Program level validation across learning curricula}\label{sec:Curriculum}

\begin{figure}[htbp]
	\centering
	\includegraphics[width=.95\textwidth]{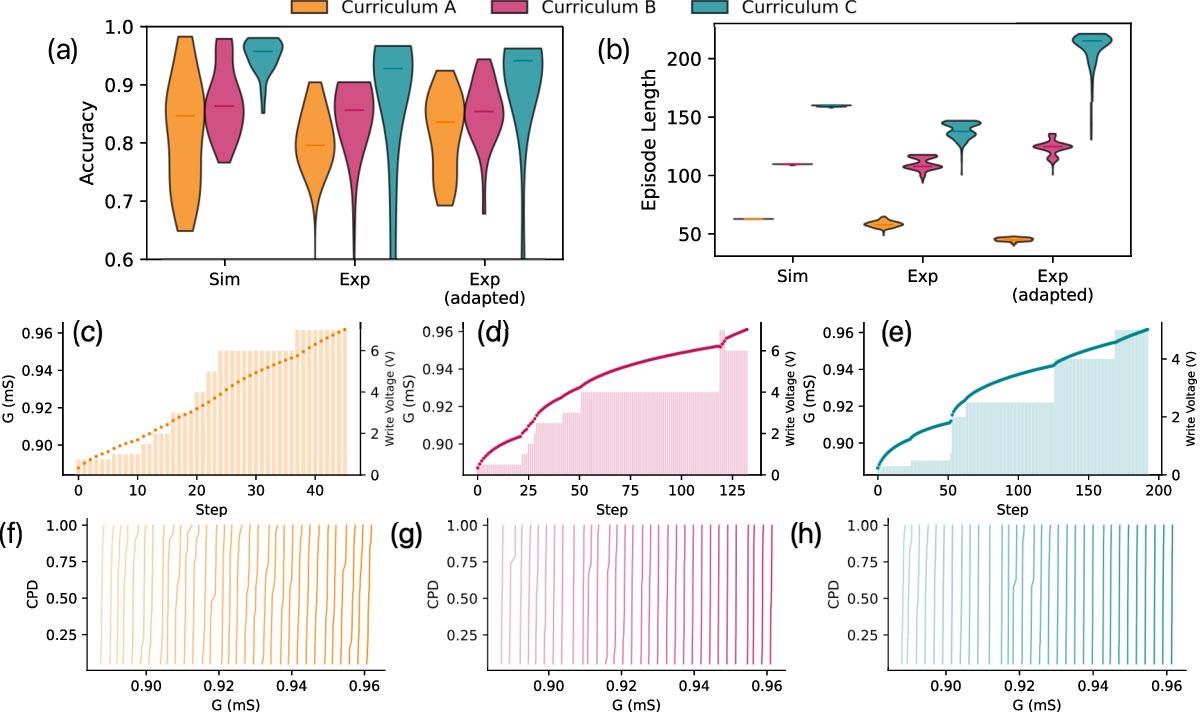}
	\caption{
(a) Grouped violin plots of programming accuracy across three curriculum configurations (Curriculum A, B, C) under three evaluation conditions: simulation (Sim), direct hardware deployment (Exp), and hardware deployment after on-device fine-tuning (Exp adapted) from 200-cycle validation. Median lines are shown within each violin. (b) Corresponding violin plots of episode length under the same three conditions. (c--e) Representative per-step conductance trajectories for Curricula A, B, and C, respectively. Scatter points (left axis) show the measured conductance evolution; bars (right axis) indicate the applied write voltage at each step. Each trajectory is selected as the median-length episode from all recorded sessions. (f--h) Conductance cumulative probability distributions (CPD) across all 32 target levels for Curricula A, B, and C, respectively. Each curve corresponds to one target level; curves with higher opacity indicate higher target conductance levels.
  }
	\label{FIG:curriculum}
\end{figure}

We evaluate the RL agent at the program level using two complementary metrics: \emph{accuracy} and \emph{episode length}. Programming accuracy quantifies the precision with which each target conductance level is reached:

\begin{equation}
    \mathrm{accuracy} = \frac{1}{L}\sum_{\ell=1}^{L}\left(1 - \frac{\overline{|G^{(\ell)} - G^*_{\ell}|}}{dG}\right),
    \label{eq:acc}
\end{equation}

\noindent where $G^{(\ell)}$ is the measured conductance when level $\ell$ is registered, $G^*_{\ell}$ is the ideal target defined in Eq.~(\ref{eq:target_levels}), and the overline is the mean over all evaluation episodes. The outer sum averages over all $L=32$ target levels, giving equal weight to precision throughout the conductance range. Episode length is the total number of write pulses applied by the agent to complete one 32-level sequence, directly measuring programming efficiency.
Each write pulse is followed by a single read operation, and the initialization of the device to the HRS before each episode is not included in this count. These metrics embody an inherent trade-off: more pulses per level improve accuracy but increase latency. The reward function provides direct control---higher $\eta_a$ favors accuracy; higher $\eta_\mathrm{step}$ or $\Delta G_\mathrm{move}$ favors efficiency---enabling different operating points to be targeted through curriculum design.

We compare three curriculum configurations whose reward schedules are summarized in Tables~\ref{tab:curriculumA}--\ref{tab:curriculumC}. Curriculum~A (used in Figs.~\ref{FIG:sim2real}--\ref{FIG:action_space}) employs aggressive step and movement penalties for efficient programming. Curriculum~B uses very small step penalties to allow long exploratory episodes. Curriculum~C applies the strictest tolerance tightening and a strong conditional movement penalty, prioritizing placement accuracy.
These schedules were designed by shifting the reward weights in the directions suggested by the trade-off described above, namely a higher $\eta_a$ and a tighter $\ell_\mathrm{tol}$ for accuracy and a larger $\eta_\mathrm{step}$ or $\Delta G_\mathrm{move}$ for efficiency.
The validity of the parameters can be verified in a simulation environment at a very low cost.
Figs.~\ref{FIG:curriculum}a and \ref{FIG:curriculum}b present violin plots of programming accuracy and episode length, respectively, under three deployment conditions (Sim, Exp, Exp adapted).
Validation was performed on a previously unseen device fabricated on the same substrate to assess the generalization capability of the learned policy across the same type of devices.
Each condition was evaluated over 200 programming episodes, enabling a statistically robust assessment of both programming performance and its cycle-to-cycle variability.
All three curricula exhibit the same qualitative accuracy pattern: a drop at direct hardware deployment that recovers after on-device fine-tuning. Post-adaptation mean accuracies are $0.82$, $0.85$, and $0.89$ for Curricula A, B, and C, respectively (Fig.~\ref{FIG:curriculum}a), consistent with each curriculum's precision emphasis.
In simulation, episode lengths are nearly deterministic, exhibiting near-zero variance even when evaluated with different random seeds.
This indicates that the fitted VTEAM surrogate captures the average device behavior but does not fully reproduce the cycle-to-cycle variability observed in the physical memristor.
After hardware deployment, the variance increases substantially, reflecting the stochastic nature of real-device switching dynamics (Fig.~\ref{FIG:curriculum}b).
Figs.~\ref{FIG:curriculum}c--e show representative conductance trajectories for Curricula A, B, and C, with the applied write-voltage sequence overlaid as bars (right axis). Distinct programming strategies emerge under different curriculum designs. Curriculum~A favors larger write voltages, resulting in larger conductance increments and rapid advancement between target states. In contrast, Curricula~B and C apply progressively more corrective actions around each target level, improving placement accuracy at the expense of programming efficiency.
Despite these differences in control strategy, all three curricula successfully maintain level separability. As shown in Figs.~\ref{FIG:curriculum}f--h, the conductance cumulative probability distributions (CPDs) of the 32 target levels remain well separated, indicating reliable multi-level programming performance. Curriculum~A achieves this with compact programming trajectories and near-uniform conductance increments, whereas Curricula~B and C produce slightly steeper CPDs through additional pulse tuning.

Taken together, these results demonstrate that the proposed framework can tailor the accuracy--efficiency characteristics of memristor programming through curriculum design. By adjusting the curriculum parameters and associated reward weighting factors, the RL agent can be guided toward different operating points along the accuracy--efficiency trade-off.
Notably, Curriculum A programs all 32 conductance levels using only 45.3 pulses on average, corresponding to approximately 1.4 write operations per level while maintaining an accuracy of 0.82.
This performance is achieved without level-specific pulse calibration, demonstrating that the RL agent can autonomously discover efficient programming policies directly from device interactions.

\begin{table}[htbp]
\centering
\caption{Reward parameter schedule for Curriculum~B.}
\label{tab:curriculumB}
\begin{tabular}{llccc}
\toprule
Stage & Mode & $\ell_\mathrm{tol}$ & $(\eta_\mathrm{step},\; \eta_a,\; \eta_\mathrm{move},\; \Delta G_\mathrm{move}\;[\mu\text{S}])$ & Max steps \\
\midrule
C0  & Simulation & 0.25 & $(0.01,\; 0.0,\; 0.5,\; 0.60)$ & 500 \\
C1  & Simulation & 0.25 & $(0.01,\; 0.2,\; 0.5,\; 0.26)$ & 400 \\
C2  & Simulation & 0.20 & $(0.01,\; 0.3,\; 0.5,\; 0.26)$ & 300 \\
C3  & Simulation & 0.15 & $(0.01,\; 0.3,\; 0.5,\; 0.26)$ & 300 \\
Exp & Hardware   & 0.15 & $(0.01,\; 0.3,\; 0.5,\; 0.26)$ & 300 \\
\bottomrule
\end{tabular}
\end{table}

\begin{table}[htbp]
\centering
\caption{Reward parameter schedule for Curriculum~C.}
\label{tab:curriculumC}
\begin{tabular}{llccc}
\toprule
Stage & Mode & $\ell_\mathrm{tol}$ & $(\eta_\mathrm{step},\; \eta_a,\; \eta_\mathrm{move},\; \Delta G_\mathrm{move}\;[\mu\text{S}])$ & Max steps \\
\midrule
C0  & Simulation & 0.15 & $(0.0,\; 0.0,\; 0.5,\; 0.24)$ & 500 \\
C1  & Simulation & 0.10 & $(0.0,\; 0.3,\; 0.5,\; 0.24)$ & 300 \\
C2  & Simulation & 0.08 & $(0.0,\; 0.4,\; 0.5,\; 0.24)$ & 300 \\
C3  & Simulation & 0.08 & $(0.0,\; 0.5,\; 0.5,\; 0.24)$ & 300 \\
Exp & Hardware   & 0.08 & $(0.0,\; 0.5,\; 0.5,\; 0.24)$ & 300 \\
\bottomrule
\end{tabular}
\end{table}

\subsection{Accuracy estimation of analog neural network}\label{sec:nn_eval}

The three RL curriculum configurations achieve post-adaptation programming accuracies of $0.82$, $0.85$, and $0.89$ (Curricula A, B, and C) with mean episode lengths of $45.3$, $124.0$, and $208.4$ write pulses, respectively.
For comparison, the FFP scheme~\cite{D4NH00623B} adapted on the same device achieves the highest programming accuracy among the evaluated methods, but requires approximately 600 write pulses for a complete programming sequence.
Together, these methods provide a range of accuracy--efficiency operating points for memristor programming, enabling the impact of programming precision on analog neural-network inference to be systematically evaluated.
Improved weight-mapping accuracy is generally expected to reduce inference errors, although the magnitude of this effect depends on the neural-network architecture and task complexity.
To quantify this impact, we estimate the neural-network inference accuracy obtained when the conductance states programmed by each method serve as analog synaptic weights.

\begin{figure}[htbp]
	\centering
	\includegraphics[width=.95\textwidth]{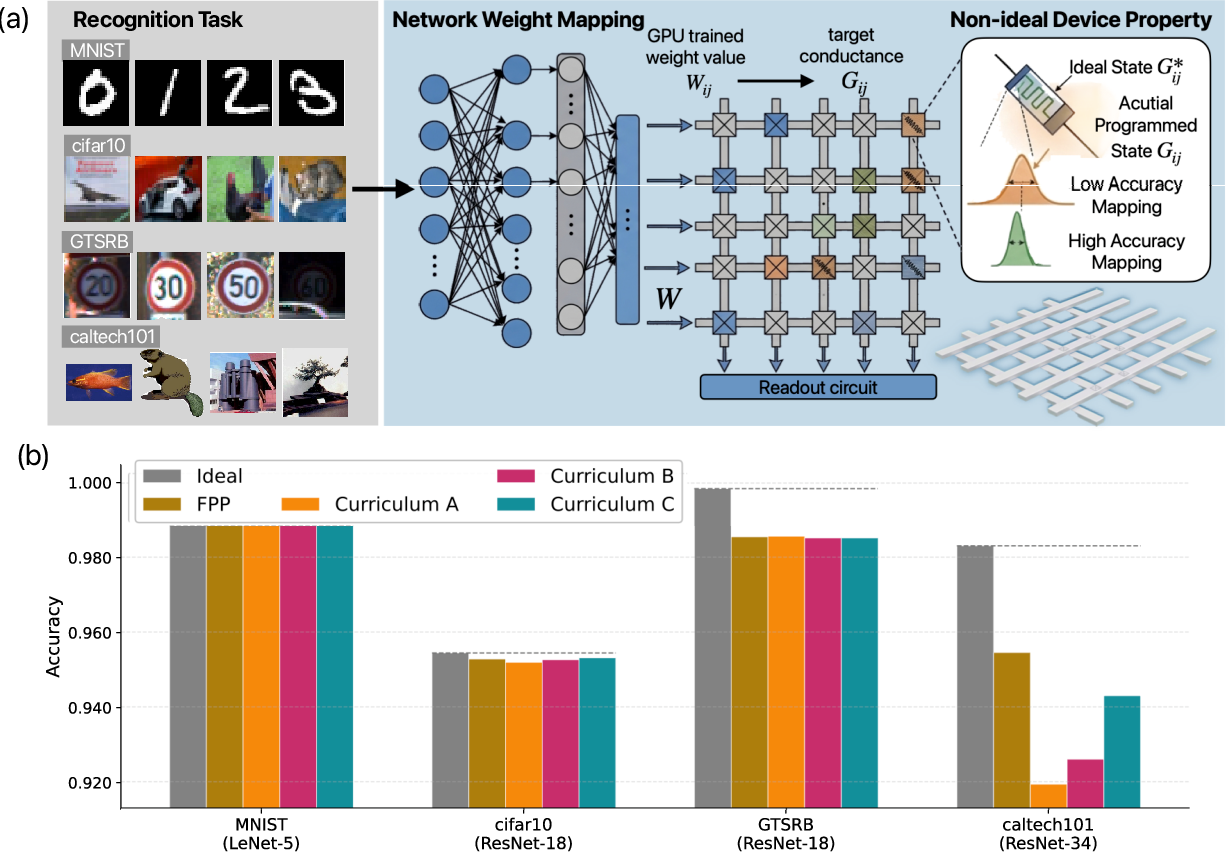}
	\caption{(a) Schematic of the image-recognition evaluation pipeline using the MNIST, CIFAR-10, GTSRB, and Caltech-101 datasets, together with the mapping of programmed memristor conductance states to neural-network weights. (b) Neural network inference accuracy achieved using conductance states programmed by five methods: Ideal (analytically perfect weight assignment), feedforward pulse (FFP) scheme~\cite{D4NH00623B}, and three RL curriculum configurations (Curriculum A, B, C). Each group of bars corresponds to a distinct inference task and model architecture. Dashed horizontal lines indicate the Ideal reference accuracy for each task.}
	\label{FIG:nn}
\end{figure}

We use the IBM Analog Hardware Acceleration Kit (AIHWKit)~\cite{9458494} to simulate inference on a resistive-memory crossbar array architecture. In a crossbar array, neural-network weights are represented by device conductances, and matrix--vector multiplication is executed in analog hardware. The inference accuracy was then evaluated by estimating the conductance-mapping error introduced when the trained network weights were mapped onto the memristor conductance states, thereby quantifying the impact of device non-idealities on analog in-memory computing performance~\cite{D4NH00623B,11496861} (Fig.~\ref{FIG:nn}a; see also Appendix~\ref{sec:crossbar} for details).

\begin{table}[htbp]
\centering
\caption{Image recognition benchmark configurations used for analog neural network inference evaluation. ($^\dagger$ means Stem modified for small-input images).}
\label{tab:nn_tasks}
{\small
\begin{tabular}{llccccc}
\toprule
Task & Backbone & Classes & Input size & Parameters & \makecell{Training\\epochs} & \makecell{Learning\\rate} \\
\midrule
MNIST       & LeNet-5             & 10  & $28\times28\times1$   & $\sim$60\,K & 30  & 1E-2  \\
CIFAR-10    & ResNet-18$^\dagger$ & 10  & $32\times32\times3$   & $\sim$11\,M & 100 & 5E-3$\rightarrow$1E-6  \\
GTSRB       & ResNet-18           & 43  & $32\times32\times3$   & $\sim$11\,M & 50  & 5E-4$\rightarrow$1E-6  \\
Caltech-101 & ResNet-34           & 101 & $224\times224\times3$ & $\sim$21\,M & 50  & 1E-4$\rightarrow$1E-6  \\
\bottomrule
\end{tabular}
}
\end{table}

We evaluate four image classification benchmarks of increasing task complexity.
MNIST is the canonical handwritten digit recognition task; CIFAR-10 and GTSRB cover natural image (10 image types) and traffic-sign classification, respectively, covering more fine-grained categories (43 sign types); and Caltech-101 spans 101 diverse object categories with high inter-class variability. Representative images from each benchmark are shown in Fig.~\ref{FIG:nn}a.
For each task, the floating-point weights of a pre-trained model are quantized to 32 conductance levels and evaluated under five conditions: (1)~\textit{Ideal}---weights analytically assigned to the nearest ideal conductance level with no device noise, representing the theoretical upper bound under 32-level quantization; (2)~\textit{FFP}---programmed using the feedforward pulse scheme ($\sim$600 pulses); and (3--5)~\textit{Curriculum A}, \textit{B}, and \textit{C}---programmed using the on-device-adapted RL policies ($45.3$, $124.0$, and $208.4$ mean pulses, respectively).
The training configurations for the ideal models using the LeNet~\cite{726791} and ResNet~\cite{He_2016_CVPR} backbones are summarized in Table~\ref{tab:nn_tasks}.

Fig.~\ref{FIG:nn}b shows the inference accuracy estimation results for these conditions.
On MNIST and CIFAR-10, the memristor-programmed platforms achieve accuracy essentially indistinguishable from the Ideal: $98.85\%$ and $\sim95\%$, respectively, with differences of less than $0.3\%$ among all five programming methods. This confirms that 32-level conductance quantization is sufficient for these tasks and that the placement-error differences among the curricula do not meaningfully impact inference quality.
On GTSRB, all five conditions converge to approximately $98\%$.
This near-identical performance across methods, despite their widely different programming accuracies, indicates that the inference bottleneck on GTSRB lies in the finite number of quantization levels rather than in placement precision.
The quantization error imposed by 32 discrete levels already limits the achievable accuracy, and further refining the placement within each level provides diminishing returns. Increasing $L$ would be the more effective path to closing the remaining gap from the Ideal.
The most sensitive benchmark is Caltech-101, where 101 output classes and a 21\,M-parameter backbone amplify the forward propagation errors. Here, accuracy degrades monotonically with programming imprecision.
In this situation, the programming conditions will influence the inference accuracy.
These results reveal a practical engineering trade-off for memristor-based analog neural network deployment.
For applications up to GTSRB complexity, Curriculum~A, which uses only 45 write pulses on average, is the most efficient choice among the evaluated strategies.
It achieves the same estimated inference quality as the much slower FFP while requiring approximately 13$\times$ fewer write pulses for weight mapping.
For more demanding tasks such as Caltech-101, the investment in additional programming pulses (Curriculum~C or FFP) yields measurable gains in accuracy.
These results demonstrate that reinforcement learning can automatically discover memristor programming policies that satisfy user-defined objectives via reward engineering, thereby enabling flexible trade-offs between programming efficiency and accuracy.
Beyond the programming task itself, this digital twin framework also establishes a fully automated workflow from device-level electrical characterization and programming optimization to simulation-based, application-level neural-network inference evaluation, bridging the gap between memristor device physics and practical edge AI applications.

\section{Discussion}\label{sec:limitations}

The central premise of this study is that the trial-and-error exploration demanded by reinforcement learning---prohibitively slow on physical memristors, where each write-and-measure cycle must be executed sequentially in real time---can be shifted almost entirely onto a physics-informed digital twin.
As a result, simulated episodes run more than four orders of magnitude faster than on-device episodes, so the bulk of policy exploration takes place in simulation and only a limited amount of on-device fine-tuning is required to recover high programming performance on real hardware.
At the same time, the analysis of the sim-to-real pipeline revealed the dependency that frames every remaining challenge: the quality of the transfer is governed by the fidelity of the digital twin. The limitations of this work, and the research directions they open, can all be read through this lens.

The most consequential limitation lies in the device model itself.
The generalized VTEAM model is a compact phenomenological surrogate rather than a complete physical description: it assumes a single internal state variable with a linear state--resistance relation, neglects thermally activated (Joule-heating-assisted) oxygen-vacancy diffusion, and represents cycle-to-cycle variability only as additive uniform noise. These simplifications surface in the calibration, where the twin is quantitatively reliable mainly for potentiation within the 1040--1130~$\Omega$ operating window and degrades at low write amplitudes ($|V|\leq0.5$~V) and for depression (overall $R^2=0.452$; Section~\ref{sec:dt_model}). Because the model was moreover calibrated only on constant-amplitude pulse trains, whereas the agent applies mixed-amplitude sequences, its accuracy for such sequences could be assessed only indirectly, through the transfer performance. Crucially, these errors are not hidden: as shown in Sections~\ref{sec:sim2real} and \ref{sec:action}, they propagate into the pretrained policy and reappear as the sim-to-real gap that the on-device fine-tuning (approximately $5\times10^{4}$ interaction steps) must close, and they are also why the RL demonstrations were confined to potentiation.
This diagnosis points directly to the way forward: enrich the twin with temperature-dependent switching kinetics and stochastic state dynamics~\cite{6226448}, validate it on mixed-amplitude sequences, and extend it to bidirectional programming.
Since the gap observed here originates from the poorly modeled low-voltage regime, improving the fidelity of the digital twin is expected to reduce the amount of on-device fine-tuning required.

A further step toward a complete demonstration of the framework's value for device development is validation at the wafer scale, where device-to-device variation becomes the dominant concern.
This variation originates in the fabrication process itself, and the limited maturity of the four-terminal planar TiO$_{2-x}$ process restricted the number and uniformity of available devices, so a systematic wafer-level study was not undertaken here.
The framework is nonetheless well suited to this regime. The workflow is not tied to any particular device, since the measurement interface, the RL environment, and the agent do not assume the switching mechanism.
Moreover, because the policy conditions each action on a window of recently observed device responses, its Transformer feature extractor affords a large receptive field~\cite{3524938.3525632} over the recent state history, allowing the agent to react to the actual state of each device.
The same policy can therefore, in principle, absorb device-to-device variation by adapting its voltage selection to different devices' responses.
These properties also make transfer to other device types attractive: once a new device is modeled and a basic set of electrical characterization data is acquired, an optimized programming strategy can be obtained directly in the simulation environment, so that the device's analog-computing potential can be explored without the extensive trial-and-error that dominates on-device switching-dynamics engineering.

More broadly, the digital-twin platform provides a natural vehicle for navigating the accuracy--cost trade-off systematically.
Because it couples a calibrated device model with an application-level inference evaluation in a single automated loop, it can be used to explore the Pareto frontier~\cite{10.1007/s10458-022-09552-y} over device types, programming strategies, and neural-network architectures, quantifying the balance between inference accuracy and programming cost for a given memristor-based AI deployment.
The closed-loop and automated nature of the framework is what makes such an exploration practical: by shifting the search into simulation and removing the manual handoffs between characterization, modeling, and evaluation, it sharply reduces the experimental cost of identifying the operating point best suited to a target application.
To support reproducibility and further development, the full system implementation is publicly released as an open-source repository~\cite{MemDigitalTwin}.

\section{Conclusion}\label{sec:conclusion}

In summary, we presented a digital twin-based framework for linear and multi-level programming of analog memristors.
The framework integrates automated device characterization, physics-based modeling, RL-based programming optimization, hardware adaptation, and simulation-based application-level neural-network evaluation into a unified closed-loop workflow.

The framework was validated on a four-terminal planar TiO$_{2-x}$ memristor and demonstrated successful 32-level linear conductance programming.
Analysis across different stages of the sim-to-real pipeline revealed that transfer performance is fundamentally limited by the fidelity of the underlying physical model, establishing a direct link between device-model accuracy and hardware deployment reliability.
In addition, curriculum-driven reward scheduling enables exploration of the programming accuracy--efficiency design space, allowing the learned policies to converge toward distinct operating points.
As a result, the framework can autonomously identify programming strategies spanning highly efficient low-pulse operation to high-precision multi-level programming, depending on the target application requirements.

The proposed methodology provides a pathway toward self-optimizing hardware systems, digital twins for device development, and autonomous materials-and-device experimentation platforms.
By establishing a closed loop between electrical characterization and physics-informed decision-making, the framework offers a scalable strategy for accelerating the development and deployment of emerging electronic devices.

%%%%%%%%%%%%%%%% REFERENCES %%%%%%%%%%%%%%%

\clearpage

\bibliography{ref}
\bibliographystyle{sciencemag}

\section*{Acknowledgments}

\paragraph*{Funding:}
This research was supported in part by a KAKENHI Grant-in-Aid (24K00926, 24K21620, 25K17654) from the Japan Society for the Promotion of Science (JSPS). This work is also partially supported by the Kyoto Technoscience Center.

\paragraph*{Author contributions:}
Z.D.: Conceptualization, methodology, software, data curation, funding acquisition, writing---original draft preparation.
T.T. and A.S.: Writing---review \& editing, funding acquisition.

\paragraph*{Competing interests:}
There are no conflicts to declare.

\paragraph*{Data and materials availability:}
The device measurement system, RL implementation, and data supporting the findings of this study are available at \url{https://github.com/DIAOZHUO/MemDigitalTwin}.

%%%%%%%%%%%%%%%% SUPPLEMENTARY INFORMATION %%%%%%%%%%%%%%%

\renewcommand{\thefigure}{S\arabic{figure}}
\renewcommand{\thetable}{S\arabic{table}}
\renewcommand{\theequation}{S\arabic{equation}}
\renewcommand{\thesection}{S\arabic{section}}
\setcounter{section}{0}
\setcounter{subsection}{0}
\setcounter{subsubsection}{0}
\setcounter{figure}{0}
\setcounter{table}{0}
\setcounter{equation}{0}
\setcounter{footnote}{0}

\clearpage
\begin{center}
{\LARGE \textbf{Supplementary Information for}}\\[0.5em]
{\Large \textbf{\scititle}}
\end{center}
\vspace{1em}

\section{Memristor device detail}\label{sec:4t}

\begin{figure}[H]
\centering
\includegraphics[width=0.65\textwidth]{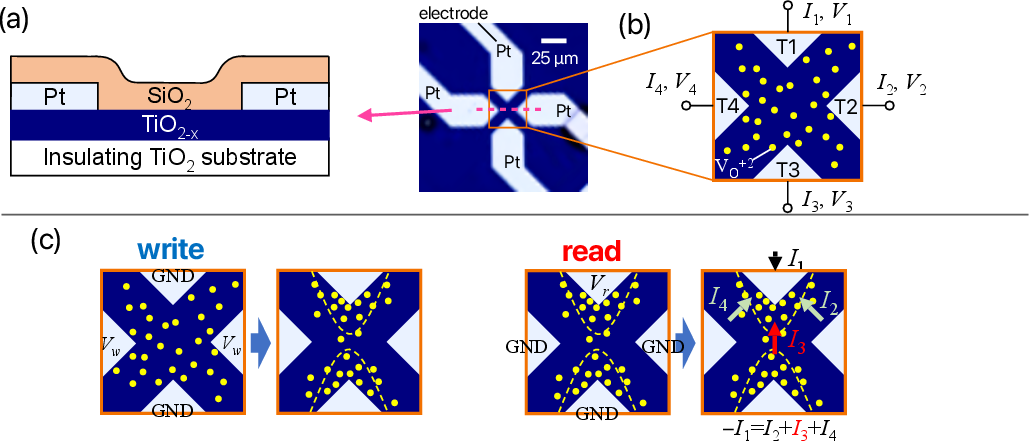}
\caption{(a) The cross-sectional diagram of the four-terminal planar TiO$_{2-x}$ memristor, followed by an optical micrograph. (b) Schematic illustration of the oxygen vacancy distribution in the resistive switching layer. (c) Write and read protocol.
The figure is reproduced from Ref.~\cite{D4NH00623B} under CC BY 3.0.}
\label{subFIG:device}
\end{figure}

The device used in this study is a four-terminal planar TiO$_{2-x}$ memristor~\cite{D4NH00623B} shown in Fig.~\ref{subFIG:device}.
The active switching layer consists of an epitaxial TiO$_{2-x}$ thin film grown on a non-doped insulating rutile TiO$_2$(001) single-crystal substrate by pulsed laser deposition, with a deposition temperature of 500~$^\circ$C and an oxygen partial pressure of approximately $10^{-4}$~Pa.
Four Pt electrodes and a SiO$_2$ passivation layer for oxidation protection are patterned by electron-beam lithography (Fig.~\ref{subFIG:device}a).

Resistive switching in this device proceeds through a bulk-type mechanism driven by the two-dimensional redistribution of positively charged oxygen vacancies (V$_\mathrm{O}^{2+}$) within the TiO$_{2-x}$ active layer (Fig.~\ref{subFIG:device}b).
Because V$_\mathrm{O}^{2+}$ carries a positive charge, an applied electric field induces drift of oxygen vacancies within the switching plane.
When vacancies accumulate in the region bridging T1 and T3, the local carrier density increases and the conductance $G_{1\text{--}3}$ rises (high-conductance state, HCS); when vacancies are depleted from this region, $G_{1\text{--}3}$ decreases (low-conductance state, LCS).
Unlike filamentary switching, in which conductance is determined by the formation or rupture of a single nanoscale conducting channel, the bulk-type mechanism distributes the conductance change across the entire active region.
This spatial averaging suppresses stochastic filament variability, resulting in smoother, more reproducible switching characteristics and enhanced cycle-to-cycle endurance.

The four-terminal geometry provides a key degree of freedom not available in conventional two-terminal devices: the electric field can be applied independently along two orthogonal axes by selecting different terminal combinations.
Voltages applied to the T2--T4 pair generate an electric field perpendicular to the T1--T3 measurement axis, enabling the V$_\mathrm{O}^{2+}$ distribution to be steered two-dimensionally within the switching plane.
This additional degree of freedom improves the controllability of the resistive switching process and enhances the reproducibility of the conductance response to programming pulses.

Fig.~\ref{subFIG:device}c shows the device write and read protocol.
To program the device conductance, a write voltage pulse $V_\mathrm{w}$ is applied simultaneously to terminals T2 and T4, while T1 and T3 are held at ground potential.
The resulting in-plane electric field drives V$_\mathrm{O}^{2+}$ redistribution along the T1--T3 axis:
a positive $V_\mathrm{w}$ concentrates vacancies between T1 and T3, increasing $G_{1\text{--}3}$ (potentiation);
a negative $V_\mathrm{w}$ depletes vacancies from this region, decreasing $G_{1\text{--}3}$ (depression).
The device conductance is measured after each write pulse by applying a small read voltage of 0.1~V to terminal T1, with T2, T3, and T4 simultaneously held at ground.
The read voltage is kept sufficiently small relative to the switching threshold to avoid perturbing the internal oxygen-vacancy distribution during measurement.

\clearpage
\section{Deep reinforcement learning network}\label{sec:nn}

\begin{figure}[H]
\centering
\includegraphics[width=\textwidth]{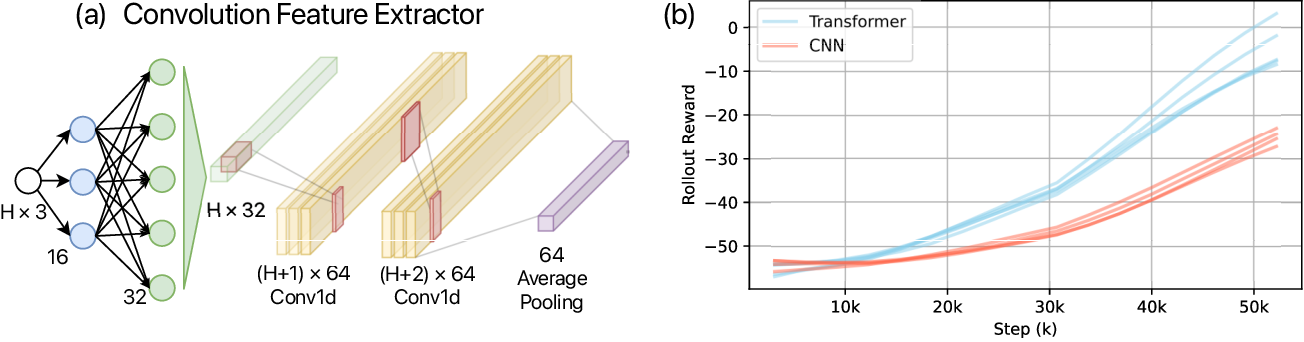}
\caption{(a) Architecture of the CNN feature extractor (see Fig.~\ref{FIG:rl_net}b for the Transformer feature extractor). (b) Rollout reward curves during Curriculum Stage 0 (C0) PPO training using Transformer and CNN feature extractors. The mean episode reward is shown for five independent training runs with observation stack counts $H$, ranging from 2 to 6. Blue curves denote the Transformer feature extractor, while red curves denote the CNN feature extractor. Curves of the same color correspond to different stack configurations.}
\label{subFIG:model}
\end{figure}

We designed two deep reinforcement learning feature extractors based on convolutional neural network (CNN) and Transformer architectures. The CNN architecture is illustrated in Fig.~\ref{subFIG:model}a, while the Transformer architecture is shown in Fig.~\ref{FIG:rl_net}b.
Here, $H$ denotes the number of stacked observations provided as the input to the feature extractor.
In the simulation environment, we evaluated both architectures with $H=2,3,4,5,6$.
The convergence efficiency was assessed by comparing the evolution of the episode reward as a function of PPO training steps, as shown in Fig.~\ref{subFIG:model}b.
Faster reward convergence indicates that the policy can be learned with fewer environment interactions, thereby reducing the sampling requirement and overall training cost.

The results show that the stack count $H$ has only a minor influence on convergence efficiency for both architectures. In contrast, the Transformer feature extractor consistently achieves faster reward growth than the CNN counterpart, indicating its superior ability to capture informative representations from the observation space.
The improved convergence suggests that the self-attention mechanism enables the Transformer to more effectively capture temporal-sequential features and long-range correlations across stacked observations than the convolution-based feature extractor, leading to more sample-efficient policy learning.
Therefore, the Transformer architecture was adopted as the feature extractor in the subsequent reinforcement learning framework.

To construct a fixed-length temporal observation sequence, the input is padded whenever fewer than ($H$) historical observations are available at the beginning of an episode.
Two padding strategies were investigated: history padding, in which the earliest available observation is repeatedly stacked until the sequence reaches length ($H$), and zero padding, in which the unavailable observations are replaced with zero vectors.
The effect of the stack length ($H = 2, 3, 4, 5, 6$) and the two observation modes was evaluated using the simulation agent after completing Curriculum Stage 3 (C3) training.

\begin{figure}[H]
\centering
\includegraphics[width=0.75\textwidth]{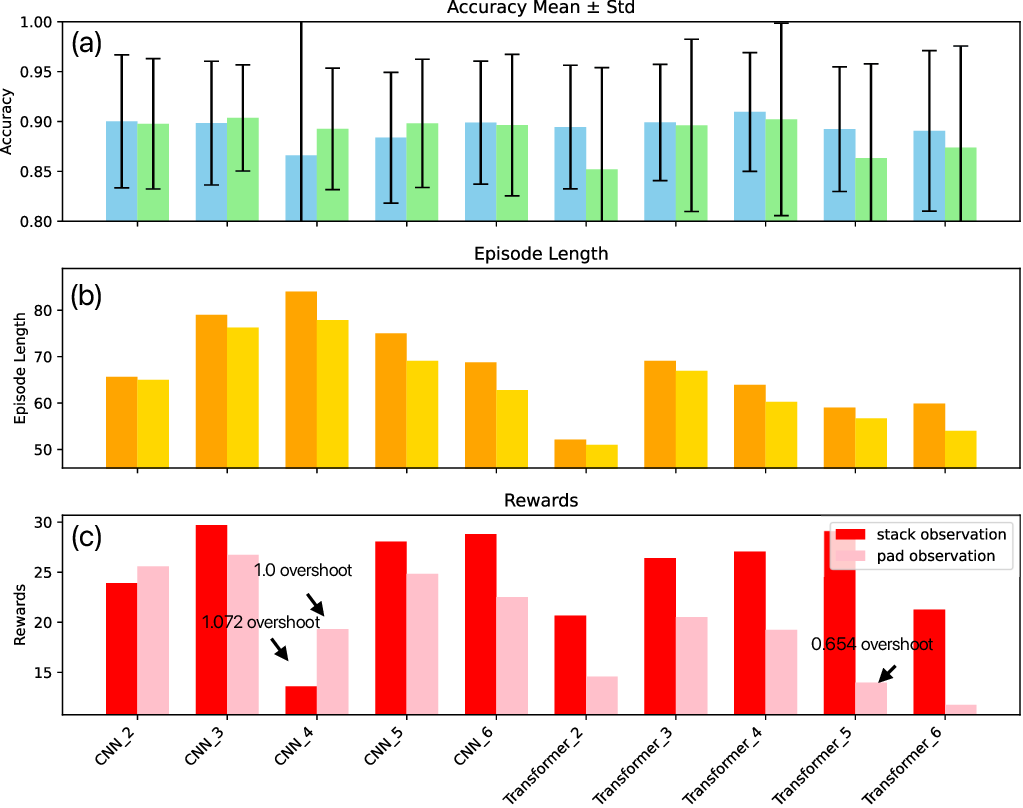}
\caption{Comparison of observation modes across feature extractor architectures. Bar charts comparing stack observation (historical observations stacked along the sequence dimension) versus pad observation (stack mode disabled; missing steps zero-padded) for CNN and Transformer feature extractors.
Three metrics are shown: programming accuracy; mean episode length; mean episodic reward.
Each pair of bars corresponds to one architecture (x-axis labels with the ``FeatureExtractor'' suffix omitted). Error caps denote one standard deviation where applicable.}
\label{subFIG:env}
\end{figure}

The trained policy was assessed in terms of (a) programming accuracy, (b) episode length, and (c) cumulative episodic reward, as summarized in Fig.~\ref{subFIG:env}.
From the perspective of programming accuracy and cumulative episodic reward, the CNN feature extractor achieves its best performance with ($H=3$), whereas the Transformer performs best with ($H=4$). Although these two configurations exhibit comparable accuracy and reward, the Transformer with ($H=4$) requires substantially fewer programming steps per episode, indicating that the agent reaches the target resistance state more efficiently. Combined with its faster convergence during PPO training, the Transformer demonstrates higher sample efficiency than the CNN architecture in both training and inference.
Based on these results, the Transformer feature extractor with a stack length of ($H=4$), together with the history-stacking observation strategy, was selected as the final deep reinforcement learning network architecture for all subsequent experiments.

\clearpage
\section{Device model fitting detail}\label{sec:fit}

\begin{figure}[H]
\centering
\includegraphics[width=\textwidth]{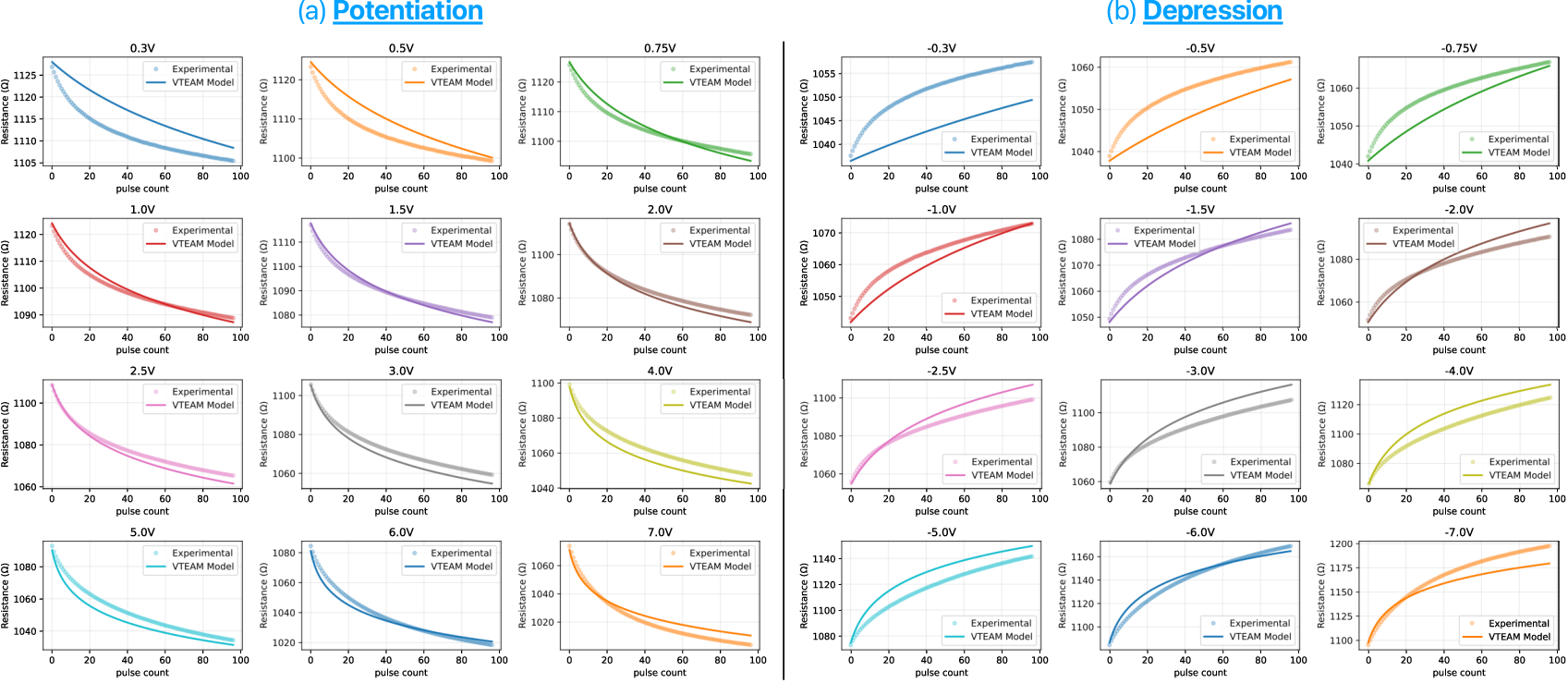}
\caption{Fitting performance of the memristor model in the proposed digital twin framework. Comparison between the model predictions and experimental measurements for 12 programming voltages during 100 consecutive (a) potentiation and (b) depression operations.}
\label{subFIG:fit}
\end{figure}

\clearpage
\section{Crossbar simulation of analog neural network}\label{sec:crossbar}

In a resistive-memory crossbar array, each synaptic weight of a neural network is encoded as the conductance of one or more memristive devices placed at the intersections of word lines and bit lines. Matrix--vector multiplication (MVM) is executed in a single analog step: an input voltage vector $\mathbf{v}$ is applied along the word lines, and the output currents accumulated along each bit line yield the dot product $\mathbf{i} = G\,\mathbf{v}$ in accordance with Ohm's and Kirchhoff's current laws, thereby performing the MVM without explicit data movement.

Because memristors store only non-negative conductance values, signed synaptic weights are represented using differential conductance pairs $(G^+,\,G^-)$, where the effective weight is proportional to $w \propto G^+ - G^-$ and both $G^+$ and $G^-$ are restricted to the device conductance range $[G_\mathrm{HRS},\, G_\mathrm{LRS}]$.
Prior to mapping, a floating-point pre-trained model is obtained by full-precision (FP32) training on GPU.
The trained weight tensors are then quantized and linearly mapped to the $L = 32$ conductance levels defined by Eq.~(\ref{eq:target_levels}), such that the full conductance span $[G_\mathrm{HRS},\, G_\mathrm{LRS}]$ covers the layer-wise weight range, normalized to the common activation output range of each hidden layer.

\begin{figure}[H]
\centering
\includegraphics[width=.75\textwidth]{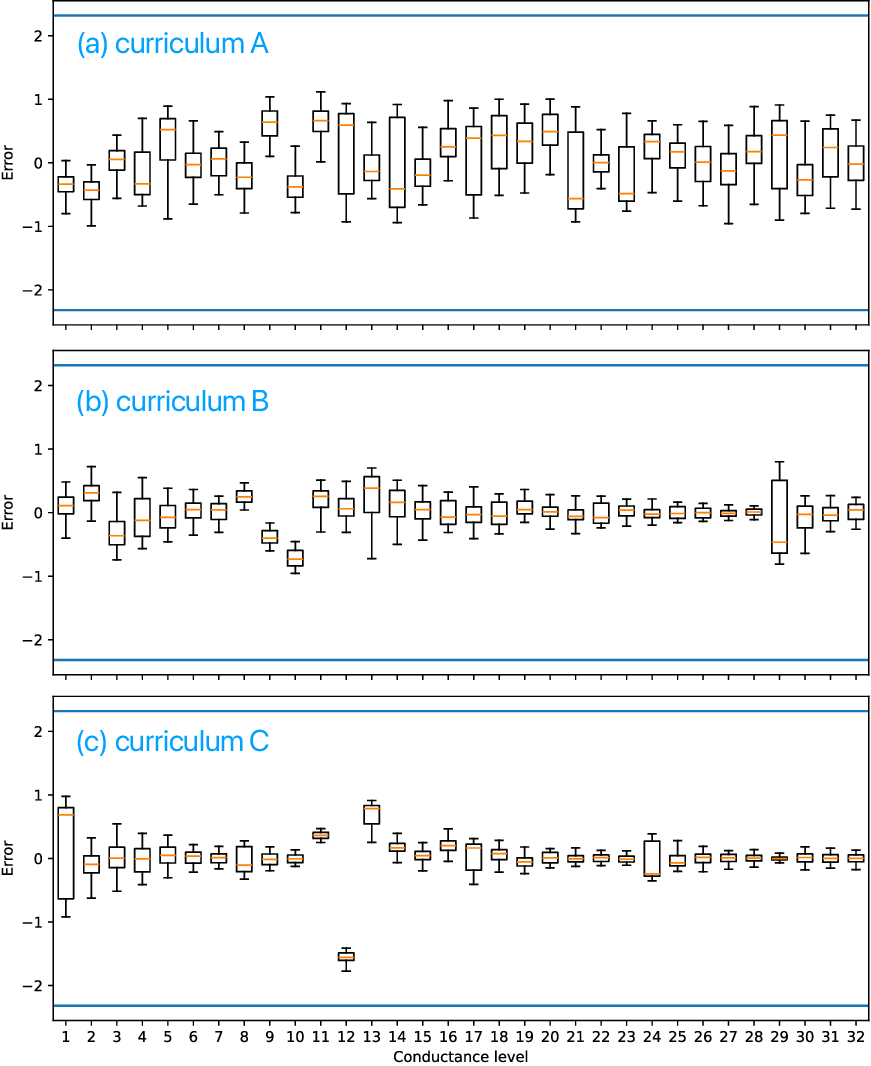}
\caption{Per-level conductance error distributions for the 3 curricula evaluated in this work.
Each subplot shows, for one of the $L=32$ target conductance levels $G^*_\ell$, the histogram of conductances actually achieved across all evaluation episodes.}
\label{subFIG:errors}
\end{figure}

Two independent sources of weight error are modeled in the crossbar inference simulation.
Placement error arises from device variability and the stochastic nature of resistive switching, which cause the conductance achieved after programming to deviate from the ideal target.
This error is characterized directly from experimental conductance measurements accumulated during hardware evaluation of each programming method.
For each of the $L = 32$ target levels $G^*_\ell$, the distribution of actually achieved conductances $\{G^{(\ell)}_k\}_{k=1}^{N}$ is recorded across all evaluation episodes ($N = 200$ per method).
The resulting per-level conductance distributions for Curricula A, B, and C are shown in Fig.~\ref{subFIG:errors}; their width directly reflects the placement accuracy defined in Eq.~(\ref{eq:acc}), with a narrower distribution indicating higher programming precision.
Quantization error arises from the finite number of discrete conductance levels ($L = 32$): any target weight that does not coincide exactly with one of the $L$ levels incurs a rounding error independent of the programming method.
This error is intrinsic to the crossbar resolution and establishes the upper bound on inference accuracy under ideal placement, corresponding to the \textit{Ideal} condition in Fig.~\ref{FIG:nn}.
During inference simulation, the weight assigned to each crossbar synapse is sampled from the corresponding per-level conductance distribution measured from the physical device.
This procedure ensures that the simulated inference accuracy reflects realistic programming imperfections rather than idealized weight placement.
In addition, AIHWKit incorporates 16-bit analog-to-digital converter (ADC) noise and analog read-circuit noise to model further practical hardware non-idealities.

\end{document}